\documentclass[11pt]{article}

\pdfoutput=1
\usepackage{arxiv}
\usepackage{hyperref}
\usepackage{url}
\usepackage[normalem]{ulem}
\usepackage{amsthm}
\usepackage{amsmath}
\usepackage{amssymb}
\usepackage{mathrsfs}
\usepackage{mathtools}
\usepackage{graphicx}
\usepackage{enumerate}
\usepackage{comment}
\usepackage{algorithm}
\usepackage{algpseudocode}
\usepackage{color}
\usepackage{stmaryrd}
\usepackage{booktabs} 
\usepackage{fullpage}
\usepackage{appendix}
\usepackage[table]{xcolor}

\theoremstyle{definition}
 
\theoremstyle{remark}

\theoremstyle{remark}

\numberwithin{theorem}{section}

\providecommand{\R}{}
\providecommand{\Z}{}
\providecommand{\N}{}

\renewcommand{\R}{\mathbb{R}}
\renewcommand{\Z}{\mathbb{Z}}
\renewcommand{\N}{{\mathbb N}}

\newcommand{\E}[1]{{\mathbb E}\left[#1\right]}										
\newcommand{\e}{{\mathbb E}}

\newcommand{\p}[1]{\mathbb{P}\left(#1\right)}

\newcommand\cD{\mathcal D}

\newcommand{\round}[1]{\ensuremath{\left\lfloor #1 \right\rceil}}

\newcommand{\tpl}[3]{\ensuremath{\mathcal{P}\left(#1,#2,#3\right)}}

\newcommand{\ABCD}{\textbf{ABCD}}
\newcommand{\ABCDoo}{\textbf{ABCD+o$^2$}}

\expandafter\def\expandafter\normalsize\expandafter{%
    \normalsize%
    \setlength\abovedisplayskip{0pt}%
    \setlength\belowdisplayskip{8pt}%
    \setlength\abovedisplayshortskip{-4pt}%
    \setlength\belowdisplayshortskip{4pt}%
}

\title{The Artificial Benchmark for Community Detection with Outliers and Overlapping Communities (\ABCDoo)}

\author{
Jordan Barrett\thanks{Department of Mathematics and Statistics, Dalhousie University, Halifax, NS, Canada; e-mail: \texttt{jbarrett@dal.ca}}
\And
Ryan DeWolfe\thanks{Department of Mathematics, Toronto Metropolitan University, Toronto, ON, Canada; e-mail: \texttt{ryan.dewolfe@torontomu.ca}}
\And
Bogumi\l{} Kami\'{n}ski\thanks{Decision Analysis and Support Unit, SGH Warsaw School of Economics, Warsaw, Poland; e-mail: \texttt{bkamins@sgh.waw.pl}} \\
\And
Pawe\l{} Pra\l{}at\thanks{Department of Mathematics, Toronto Metropolitan University, Toronto, ON, Canada; e-mail: \texttt{pralat@torontomu.ca}}
\And
Aaron Smith\thanks{Department of Mathematics and Statistics, University of Ottawa, Ottawa, ON, Canada; e-mail: \texttt{asmi28@uOttawa.ca}}
\And
Fran\c{c}ois Th\'{e}berge\thanks{Tutte Institute for Mathematics and Computing, Ottawa, ON, Canada; email: \texttt{theberge@ieee.org}}
}

\begin{document}
    
\maketitle

\begin{abstract}
The \textbf{A}rtificial \textbf{B}enchmark for \textbf{C}ommunity \textbf{D}etection (\textbf{ABCD}) graph is a random graph model with community structure and power-law distribution for both degrees and community sizes. The model generates graphs similar to the well-known \textbf{LFR} model but it is faster, more interpretable, and can be investigated analytically.
In this paper, we use the underlying ingredients of the \textbf{ABCD} model, and its generalization to include outliers (\textbf{ABCD+o}), and introduce another variant that allows for overlapping communities, \ABCDoo. 
\end{abstract}
\keywords{
ABCD, community detection, overlapping communities, benchmark models
}
%
%
%

%%%%%%%%%%%%%%%%%%%%%%%%%%%%%%%%%%%%%%%%%%%%%%%%%%%%%%%%%%%
\section{Introduction}\label{sec:intro} 
%%%%%%%%%%%%%%%%%%%%%%%%%%%%%%%%%%%%%%%%%%%%%%%%%%%%%%%%%%%

One of the most important features of real-world networks is their community structure, as it reveals the internal organization of nodes. In social networks, communities may represent groups by interest; in citation networks, they correspond to related papers; in the Web graph, communities are formed by pages on related topics, etc. Identifying communities in a network is therefore valuable, as it helps us understand the structure of the network.

Detecting communities is quite a challenging task. In fact, there is no definition of community that researchers and practitioners agree on. Still, it is widely accepted that a community should induce a graph that is denser than the global density of the network~\cite{girvan2002community}. Numerous community detection algorithms have been developed over the years, using various techniques such as optimizing modularity, removing high-betweenness edges, detecting dense subgraphs, and statistical inference. We direct the interested reader to the survey~\cite{fortunato2010community} or one of the numerous books on network science~\cite{kaminski2021mining}.

Most community detection algorithms aim to find a partition of the set of nodes, that is, a collection of pairwise disjoint communities with the property that each node belongs to exactly one of them. This is a natural assumption for many scenarios. For example, most of the employees on LinkedIn work for a single employer. On the other hand, users of Instagram can belong to many social groups associated with their workplace, friends, sports, etc. Researchers might be part of many research groups. A large fraction of proteins belong to several protein complexes simultaneously. As a result, many real-world networks are better modelled as a collection of overlapping communities~\cite{palla2005uncovering} and, moreover, many community detection problems should be tackled by finding overlapping sets of nodes rather than a partition. 

In the context of overlapping communities, detection is even more challenging. For example, in non-overlapping community detection one can easily verify that a node $i$ is highly connected to its proposed community $C$, whereas in overlapping community detection a node $i$ might be assigned to many communities $C_1, \dots, C_k$, and the number of connections from $i$ into each community may vary drastically.

To support the development and analysis of overlapping community detection algorithms, a large and diverse catalogue of networks with ground-truth communities is required for testing, tuning and training. Unfortunately, there are few such datasets with communities properly identified and labelled. As a result, there is a need for synthetic random graph models with community structure that resemble real-world networks to benchmark and tune unsupervised clustering algorithms. The popular \textbf{LFR} (\textbf{L}ancichinetti, \textbf{F}ortunato, \textbf{R}adicchi) model~\cite{lancichinetti2008benchmark,lancichinetti2009benchmarks} generates networks with communities and, at the same time, allows for heterogeneity in the distributions of both node degrees and of community sizes. Due to this structural freedom, the \textbf{LFR} model became a standard and extensively used model for generating artificial networks with ground-truth communities. 

A model similar to \textbf{LFR}, the \textbf{A}rtificial \textbf{B}enchmark for \textbf{C}ommunity \textbf{D}etection (\textbf{ABCD})~\cite{kaminski2021artificial}, was recently introduced and implemented\footnote{\url{https://github.com/bkamins/ABCDGraphGenerator.jl/}}, along with a faster and multithreaded implementation\footnote{\url{https://github.com/tolcz/ABCDeGraphGenerator.jl/}} (\textbf{ABCDe})~\cite{kaminski2022properties}. Undirected variants of \textbf{LFR} and \textbf{ABCD} produce graphs with comparable properties, but \textbf{ABCD} (and especially \textbf{ABCDe}) is faster than \textbf{LFR} and can be easily tuned to allow the user to make a smooth transition between the two extremes: pure (disjoint) communities and random graphs with no community structure. Moreover, \textbf{ABCD} is easier to analyze theoretically---for example, in~\cite{kaminski2022modularity} various properties of the \ABCD{} model are investigated, including the modularity which, despite some known issues such as the ``resolution limit'' reported in~\cite{fortunato2007resolution}, is an important graph property in the context of community detection. In~\cite{self-similarityABCD}, some interesting and desired self-similar behaviour of the \textbf{ABCD} model is discovered; namely, that the degree distribution of ground-truth communities is asymptotically the same as the degree distribution of the whole graph (appropriately normalized based on their sizes). Finally, the building blocks in the model are flexible and may be adjusted to satisfy different needs. Indeed, the original \textbf{ABCD} model was adjusted to include outliers (\textbf{ABCD+o})~\cite{kaminski2023artificial} and was generalized to hypergraphs (\textbf{h--ABCD})~\cite{kaminski2023hypergraph}\footnote{\url{https://github.com/bkamins/ABCDHypergraphGenerator.jl}}. For these reasons \textbf{ABCD} is gaining recognition as a benchmark for community detection algorithms. For example, in~\cite{aref2022bayan} the authors use both \textbf{ABCD} and \textbf{LFR} graphs to compare 30 community detection algorithms and, in their work, highlight that \emph{``being directly comparable to \textbf{LFR}, \textbf{ABCD} offers additional benefits, including higher scalability and better control for adjusting an analogous mixing parameter.''}

In this paper we introduce the \ABCDoo\ model: a generalization of the \textbf{ABCD+o} model that allows for overlapping communities. The \textbf{LFR} model has been extended in a similar way~\cite{lancichinetti2009benchmarks}, and in this model the nodes are assigned to communities based on the construction of a random bipartite graph between nodes and communities resulting in (a) a small amount of overlap between almost every pair of communities, and (b) rarely any pair of communities with a large overlap. In \ABCDoo, we instead generate overlapping communities based on a hidden, low-dimensional geometric layer which tends to yield fewer and larger overlaps. This geometric approach to community structure is justified by the fact that latent geometric spaces are believed to shape complex networks (e.g., social media networks shaped by users' opinions, education, knowledge, interests, etc.). These latent spaces have been successfully employed for many years to model and explain network properties such as self-similarity~\cite{serrano2008self}, homophily and aversion~\cite{henry2011emergence}. For more details, we direct the reader to the survey~\cite{boguna2021network} or the book~\cite{serrano2021shortest}. In addition to the geometric layer, the ancillary benefits of the \textbf{ABCD} model (an intuitive noise parameter, a fast implementation, and theoretical analysis) are still present, making the \ABCDoo\ model an attractive option for benchmarking community detection algorithms.

As mentioned above, modelling complex networks using underlying geometric hidden layers is not new. However, in most spatial models such as the Spatial Preferential Attachment (SPA) model~\cite{aiello2008spatial}, nodes are embedded in a metric space, and link formation is influenced by the metric distance between vertices. The metric space is meant to be a feature space, where coordinates represent information associated with the node. For example, in text mining, documents are commonly represented as vectors in a word space. The metric is chosen so that metric distance represents similarity, i.e.\ nodes whose information entities are closely related will be at a short distance from each other in the metric space. In the \ABCDoo\ model, the underlying geometric space plays a slightly different role. In this case, the geometric layer affects the formation of communities which group together nodes that are close to each other. For example, communities in social networks tend to group together users with similar interests, age, geographical locations, all of which can be represented as vectors in a metric space. This natural mechanism creates correlated overlapping clusters which might be challenging to obtain via other means. Connections between users in such social networks are formed as a consequence of such underlying overlapping structure. We are not aware of similar approach used in other models but, given it is quite natural, we would not be surprised if similar techniques were already used.

\bigskip

The remainder of the paper is organized as follows. In Section~\ref{sec:ABCDoo} we present the \ABCDoo{} model, with a full description of generating a graph in Section~\ref{subsec:abcdoo construction}. Next, in Section~\ref{sec:properties of model} we show some properties of the model and compare these properties to those of real graphs with known overlapping communities. In Section~\ref{sec:benchmarking} we demonstrate an application of the model by benchmarking community detection algorithms and comparing their quality under different levels of noise and overlap. Finally, some concluding remarks are given in Section~\ref{sec:conclusion}.

%%%%%%%%%%%%%%%%%%%%%%%%%%%%%%%%%%%%%%%%%%%%%%%%%%%%%%%%%%%
\section{\ABCDoo\ --- \ABCD\ with \textbf{O}verlapping Communities and \textbf{O}utliers}\label{sec:ABCDoo}
%%%%%%%%%%%%%%%%%%%%%%%%%%%%%%%%%%%%%%%%%%%%%%%%%%%%%%%%%%%

As mentioned in the introduction, the original \ABCD\ model was extended to include outliers resulting in the \textbf{ABCD+o} model. For our current needs, we extend \textbf{ABCD+o} further to allow for non-outlier nodes to belong to multiple communities, resulting in the \ABCDoo\ model, \ABCD\ with \textbf{o}verlapping communities and \textbf{o}utliers. 

%%%%%%%%%%%%%%%%%%%%%%%%%%%%%%%%%%%%%%%%%%%%%%%%%%%%%%%%%%%
\subsection{Notation} 

For a given $n \in \N := \{1, 2, \ldots \}$, we use $[n]$ to denote the set consisting of the first $n$ natural numbers, that is, $[n] := \{1, 2, \ldots, n\}$. 

We use standard probability notation throughout the paper. For a random variable $X$, write $\p{X = k}$ for the probability that $X = k$, and write $\E{X}$ for the expected value of $X$. For a distribution $\cD$, write $X \sim \cD$ to mean $X$ is sampled from the distribution $\cD$. 

For any real number $x = a+b$ with $a \in \Z$ and $b \in [0,1)$, the random variable $\round{x}$ is defined as

\[
\round{x} := \bigg\{
\begin{array}{ll}
a & \text{ with probability } 1-b, \text{ and}\\
a+1 & \text{ with probability } b \,.
\end{array}
\]
Note that $\E { \round{x} } = a (1-b) + (a+1) b = x$. 

Power-law distributions will be used to generate both the degree sequence and community sizes so let us formally define it. For given parameters $\gamma \in (0, \infty)$, $\delta, \Delta \in \N$ with $\delta \leq \Delta$, we define a truncated power-law distribution $\tpl{\gamma}{\delta}{\Delta}$ as follows. For $X \sim \tpl{\gamma}{\delta}{\Delta}$ and for $k \in \N$ with $\delta \leq k \leq \Delta$,

\begin{equation}\label{eq:tpl}
\p{X = k} = \frac{\int_k^{k+1} x^{-\gamma} \, dx}{\int_{\delta}^{\Delta+1} x^{-\gamma} \, dx} \,.
\end{equation}

%%%%%%%%%%%%%%%%%%%%%%%%%%%%%%%%%%%%%%%%%%%%%%%%%%%%%%%%%%%
\subsection{The Configuration Model}\label{subsec:config}

The well-known configuration model is an important ingredient of all variants of the \textbf{ABCD} models, so let us formally define it here. Suppose that our goal is to create a graph on $n$ nodes with a given degree sequence $\textbf{d} := (d_i, i \in [n])$, where $\textbf{d}$ is a sequence of non-negative integers such that $m := \sum_{i \in [n]} d_i$ is even. We define a random multi-graph $\mathrm{CM}(\textbf{d})$ with a given degree sequence known as the \textbf{configuration model} (sometimes called the \textbf{pairing model}), which was first introduced by Bollob\'as~\cite{bollobas1980probabilistic}. (See~\cite{bender1978asymptotic,wormald1984generating,wormald1999models} for related models and results.)

We start by labelling nodes as $[n]$ and, for each $i \in [n]$, endowing node $i$ with $d_i$ half-edges. We then iteratively choose two unpaired half-edges uniformly at random (from the set of pairs of remaining half-edges) and pair them together to form an edge. We iterate until all half-edges have been paired. This process yields a graph $G_n \sim \mathrm{CM}(\textbf{d})$ on $n$ nodes, where $G_n$ is allowed self-loops and multi-edges and thus $G_n$ is a multi-graph.

%%%%%%%%%%%%%%%%%%%%%%%%%%%%%%%%%%%%%%%%%%%%%%%%%%%%%%%%%%%
\subsection{Parameters of the \ABCDoo\ Model}

Table~\ref{tab:parameters} summarizes the parameters that govern the \ABCDoo\ model. Note that the ranges for parameters $\gamma$ and $\beta$ are suggestions chosen according to experimental values commonly observed in complex networks~\cite{barabasi2016network,orman2009comparison}. In fact, users may inject any degree sequence and sequence of community sizes as inputs to the model.

\begin{table}
\[
\begin{array}{|l|l|l|}
\hline
\text{Parameter} & \text{Range} & \text{Description}\\
\hline
n & \N & \text{Number of nodes} \\
s_0 & \N & \text{Number of outliers} \\
\eta & [1,\infty) & \text{Average number of communities a non-outlier node is part of } \\
d & \N & \text{Dimension of reference layer} \\
\rho & [-1,1] & \text{Pearson correlation between the degree of nodes and} \\
& & \text{the number of communities they belong to} \\
\hline
\gamma & (2,3) & \text{Power-law degree distribution with exponent } \gamma\\
\delta & \N & \text{Min degree as least } \delta \\
\Delta & \{ \delta, \delta+1, \ldots \} & \text{Max degree at most } \Delta \\
\hline
\beta & (1,2) & \text{Power-law community size distribution with exponent } \beta\\
s & \{ \delta+1, \delta+2, \ldots \} & \text{Min community size at least } s \\
S & \{ s, s+1, \ldots \} & \text{Max community size at most } S \\
\hline
\xi & [0,1] & \text{Level of noise}\\
\hline
\end{array}
\]
\caption{Parameters of the \ABCDoo\ model.\label{tab:parameters}}
\end{table}

%%%%%%%%%%%%%%%%%%%%%%%%%%%%%%%%%%%%%%%%%%%%%%%%%%%%%%%%%%%
\subsection{Big Picture}

The \ABCDoo\ model generates a random graph on $n$ nodes with a degree sequence and a community size sequence that follow truncated power laws with exponents $\gamma$ and, respectively, $\beta$. The degree sequence is labelled as $(d_i, i \in [n])$ and the community size sequence is labelled as $(s_j, j \in [L])$, and note that $L$ is a random variable. The nodes are split into $s_0$ outliers and $\hat{n}=n-s_0$ non-outliers. The non-outliers span a collection of $L$ communities $(C_j,j \in [L])$, with $|C_j| = s_j$ and with each non-outlier belonging to at least one community. These communities are generated, and will overlap (unless $\eta = 1$), according to a hidden $d$-dimensional reference layer in such a way that non-outliers will belong to $\eta$ communities, on average. The model allows for a correlation $\rho$ between the degree of a node and the number of communities it belongs to. In some cases, though, a correlation of $\rho$ is unachievable in which case we achieve the highest correlation possible.

Parameter $\xi \in [0,1]$ dictates the amount of noise in the network, i.e., the fraction of edges that do not lie in a single community. Each non-outlier node $i$ has its degree $d_i$ split into two parts: \emph{community degree} $Y_i$ and \emph{background degree} $Z_i$ (and thus $d_i=Y_i+Z_i$). Using a random rounding technique, we choose the community degrees and background degrees so that $\E{Y_i} = (1-\xi) d_i$ and $\E{Z_i} = \xi d_i$. Note that the neighbours of outliers are sampled from the entire graph, ignoring the underlying community structure, meaning $Y_i = 0$ and $Z_i = d_i$ if $i$ is an outlier. 

Once nodes are assigned to communities and their degrees are split, the edges of each community graph are independently generated by the configuration model on the corresponding community degree sequences, i.e., community graph $G_j$ corresponding to community $C_j$ is generated by the configuration on the degree sequence $(Y_i, i \in C_j)$. Once communities are generated, the background graph is generated by the configuration model on the degree sequence $(Z_i, i \in [n])$. The final \ABCDoo\ model, after an additional clean-up phase to rewire self-loops and duplicate edges, is the union of the community graphs and the background graph.

\subsection{The \ABCDoo\ Construction}\label{subsec:abcdoo construction}

The following 6-phase construction process generates the \ABCDoo\ synthetic networks. 

\subsubsection*{Phase 1: creating the degree distribution.}

This phase is the same as in the original \ABCD\ model and its generalization, \textbf{ABCD+o}. By default, we sample $d_i \sim \tpl{\gamma}{\delta}{\Delta}$, independently for each $i \in [n]$, then re-label the samples so that $d_1 \geq \dots \geq d_n$. To ensure that $\sum_{i \in [n]} d_i$ is even, we decrease $d_1$ by 1 if necessary; we relabel again if needed to ensure that the sequence remains monotone. Alternatively, the degree sequence can be given explicitly as an input.

\subsubsection*{Phase 2: assigning nodes as outliers.}

This phase is also the same as in the \textbf{ABCD+o} model. As mentioned in the big picture summary, the neighbours of outliers will be sampled from the entire graph, ignoring the underlying community structure. It feels that this part is straightforward, but a problem might occur when $\xi$ is close to zero. In the extreme case, when $\xi=0$, the edges in the background graph are precisely the edges connecting two outliers, and for a simple background graph to exist all outliers must have degree at most $s_0 - 1$. To handle the more general issue that arises when $\xi$ is small, we need to estimate the number of nodes with at least one edge in the background graph, then use this estimation to upper-bound the degree of an outlier.

The \textit{background degree} of a node is the number of edges in the background graph containing said node. The expected background degree of node $i$ of degree $d_i$ is $\round {\xi d_i}$ (recall the definition of $\round{\cdot}$ given in Section~2.1; more details will be provided in Phase~5). In particular, the background degree of node $i$ is positive with probability $\min(1, \xi d_i)$. Hence, before it is decided which nodes become outliers, we expect $\ell = \sum_{i\in [n]}\min(1, \xi d_i)$ nodes with positive background degree. Moreover, the $s_0$ outliers are guaranteed to have positive background degree. Thus, in total we expect $\ell + (n-\ell) (s_0 / n)$ nodes with positive background degree. Therefore, we insist that a node $i$ of degree $d_i$ cannot become an outlier unless
\begin{equation}\label{eq:outliers}
d_i \le \ell + s_0 - \ell s_0/n -1.
\end{equation}
In practice (when the number of nodes $n$ is large, the number of outliers $s_0$ is relatively small, and the level of noise $\xi$ is not zero), there are plenty of nodes with a non-zero degree in the background graph, and so there is no restriction needed for outliers. Nevertheless, we include the restriction to ensure no issues during Phase~6; a subset of $s_0$ nodes satisfying~(\ref{eq:outliers}) are selected uniformly at random to become outliers. 

\subsubsection*{Phase 3: creating overlapping communities.}

By the end of Phase~2, we have a degree sequence $(d_i, i \in [n])$ and an assignment of degrees to outliers and non-outliers. It is important to keep in mind that, although communities are created in this phase, we do not delegate degrees to these communities until Phase~4. We instead construct a collection of \textit{elements} which will be points in $d$-dimensional space, each assigned to some number of communities. We will construct $\hat{n} = n-s_0$ elements and they will belong to $\eta \ge 1$ communities on average. To be compatible with the original \ABCD\ model, each element will belong to a single \textit{primary} community, and these primary memberships will partition the elements. Then, we will grow each community by a factor of $\eta$, adding nearby elements in the $d$-dimensional space, so that the collective size of all communities is equal (in expectation) to $\eta \hat{n} = \eta (n-s_0)$ (we define this growth process formally below). For each element $v$ added to community $C$ during the growth process, $C$ will be referred to as a \textit{secondary} community of $v$.

Similar to the degree sequence, the community size sequence $(s_j, j \in [L])$ follows a power-law with parameter $\beta$, minimum value $s$, and maximum value $S$. Hence, the sequence of primary communities $(\hat{s}_j, j \in [L])$ will need to follow a power-law with parameter $\beta$, minimum value $\hat{s} = \lceil s/\eta \rceil$, and maximum value $\hat{S} = \lfloor S/\eta \rfloor$. In addition, we require $\sum_{j \in [L]} \hat{s}_j = \hat{n}$. To satisfy both requirements, we sample $\hat{s}_j \sim \tpl{\beta}{\hat{s}}{\hat{S}}$ independently until the sum of samples is at least $\hat{n}$. If, at this point, the sum is $\hat{n} + a$ with $a > 0$ then we perform one of two actions. If the last added sample has size at least $a+\hat{s}$, we reduce it by $a$. Otherwise (that is, if the last added sample has size $c < a+\hat{s}$), we delete this sample, select $c-a$ old samples uniformly at random and increase each by 1. Lastly, let $L$ be the random variable counting the number of communities, and relabel $(\hat{s}_j, j \in [L])$ so that $\hat{s}_1 \geq \dots \geq \hat{s}_L$. Similar to the degree sequence, the sequence of community sizes can instead be given explicitly as an input. Each primary community of size $\hat{s}_j$ will grow to size $s_j = \round{ \eta \hat{s}_j}$. As a result, 

$$
\E{\sum_{j \in [L]} s_j} 
= \sum_{j \in [L]} \E{s_j} 
= \sum_{j \in [L]} \E{\round{\eta \hat{s}_j}}
= \eta \sum_{j \in [L]} \hat{s}_j 
= \eta \hat{n} 
= \eta (n-s_0),
$$
as desired. 

We now create our $d$-dimensional reference layer that will guide the process of assigning elements to communities. One may think of this reference layer as a map representing various latent properties of elements (such as people’s age, education, geographic location, beliefs, etc.) shaping communities (such as communities in social media) based on proximity. In this reference layer, each of the $\hat{n}$ elements is generated as a vector in $\R^d$, sampled independently and uniformly at random from the ball of radius $1$ centred at the origin $\textbf{0} = (0,0,\ldots,0)$. Let $R$ be the set of $\hat{n}$ elements and initalize $R_1 = R$. We assign elements to their primary communities, dealing with one primary community at a time. To form primary community $\hat{C}_j$, we first select the element in $R_j$ furthest from $\textbf{0}$. This element, together with its $\hat{s}_j-1$ nearest neighbours in $R_j$, form the elements of $\hat{C}_j$. We then remove these elements from $R_j$ to construct $R_{j+1}$ and repeat the process until all elements have been assigned a primary community. At the end of the process, each element belongs to exactly one primary community and the collection $(\hat{C}_j, j \in [L])$ partitions the elements.

We are now ready to define the growth process as mentioned above. We grow each primary community $\hat{C}_j$ of size $\hat{s}_j$ to form the full community $C_j$ of size $s_j$. Initially, we set $C_j = \hat{C}_j$ for all $j \in [L]$. Each community will grow independently, meaning we can grow them in any order (or in parallel). As before, let $R$ be the set of $\hat{n}$ elements. For primary community $\hat{C}_j$, let $\mathbf{x}_j \in \R^d$ be the center of mass (mean) of $\hat{C}_j$. We investigate elements of $R$ in order of increasing distance from $\mathbf{x}_j$. During the investigation, if an element $v \in R$ is not a member of $\hat{C}_j$, then we assign $C_j$ as a secondary community of $v$. We stop investigating once the number of members in $C_j$ (both primary and secondary) is $s_j$. Write $\eta_v$ for the number of communities containing element $v \in R$.

\begin{figure}[H]
    \centering
    \includegraphics[width=0.32\linewidth]{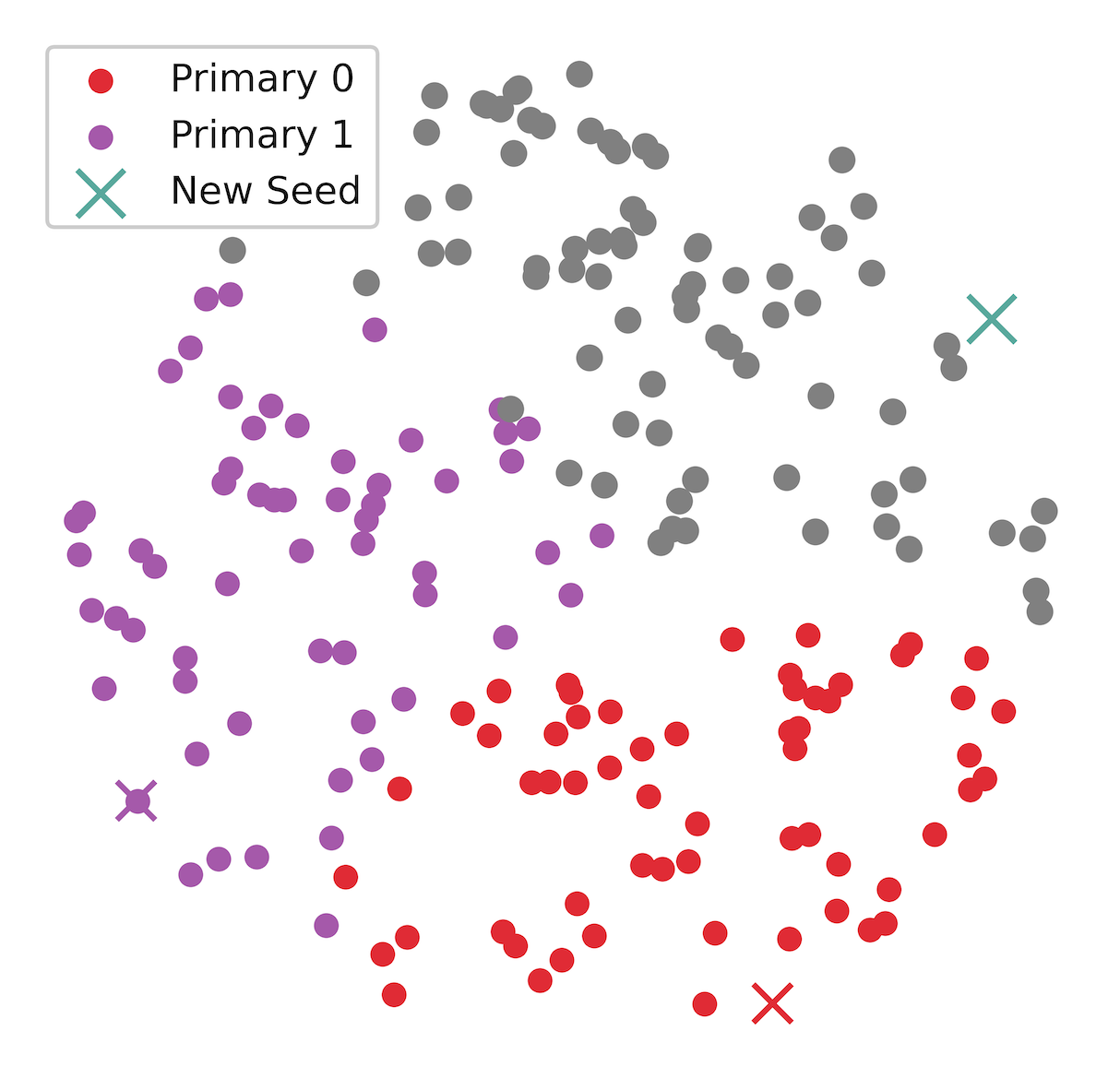}
    \includegraphics[width=0.32\linewidth]{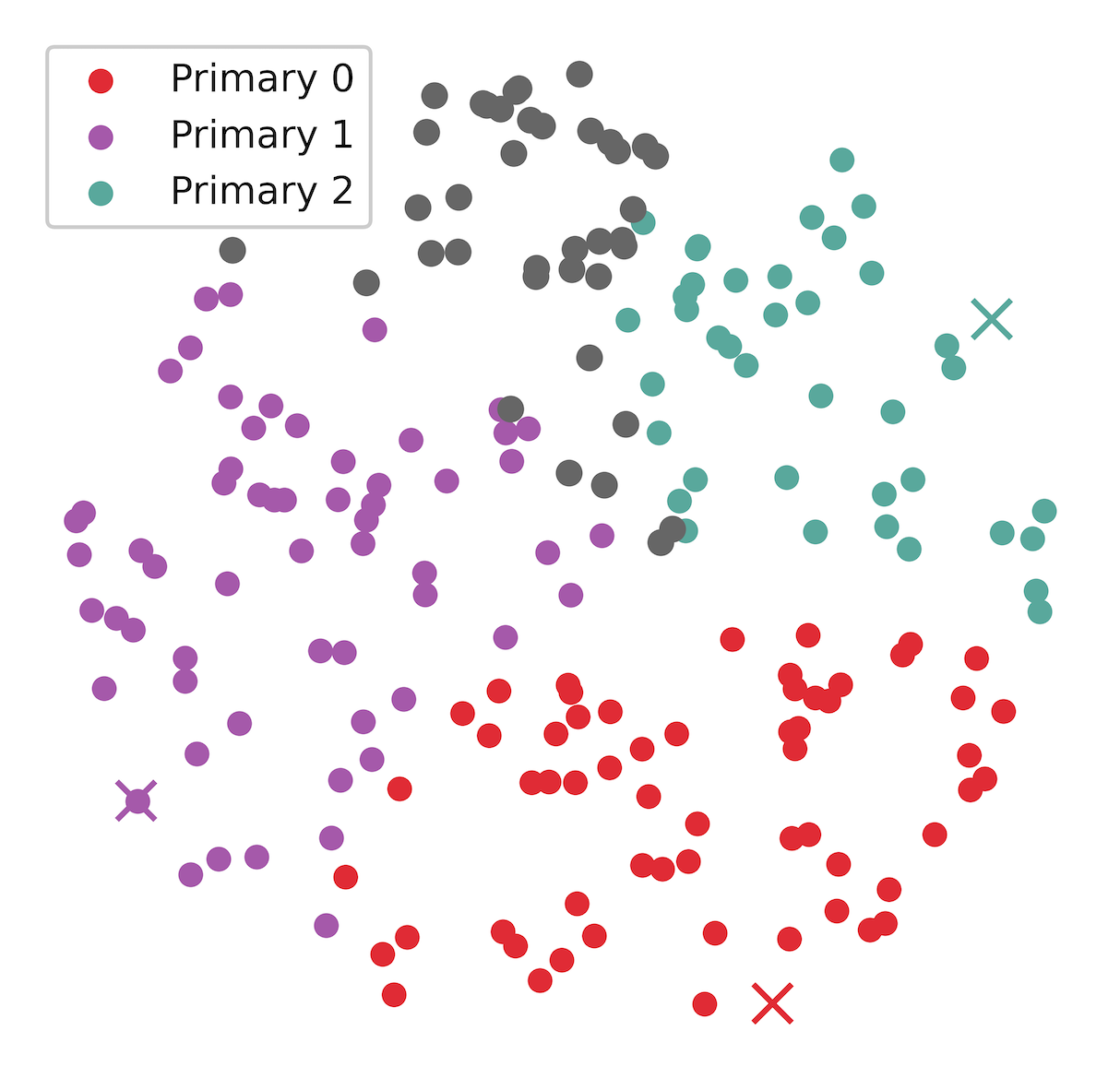}
    \includegraphics[width=0.32\linewidth]{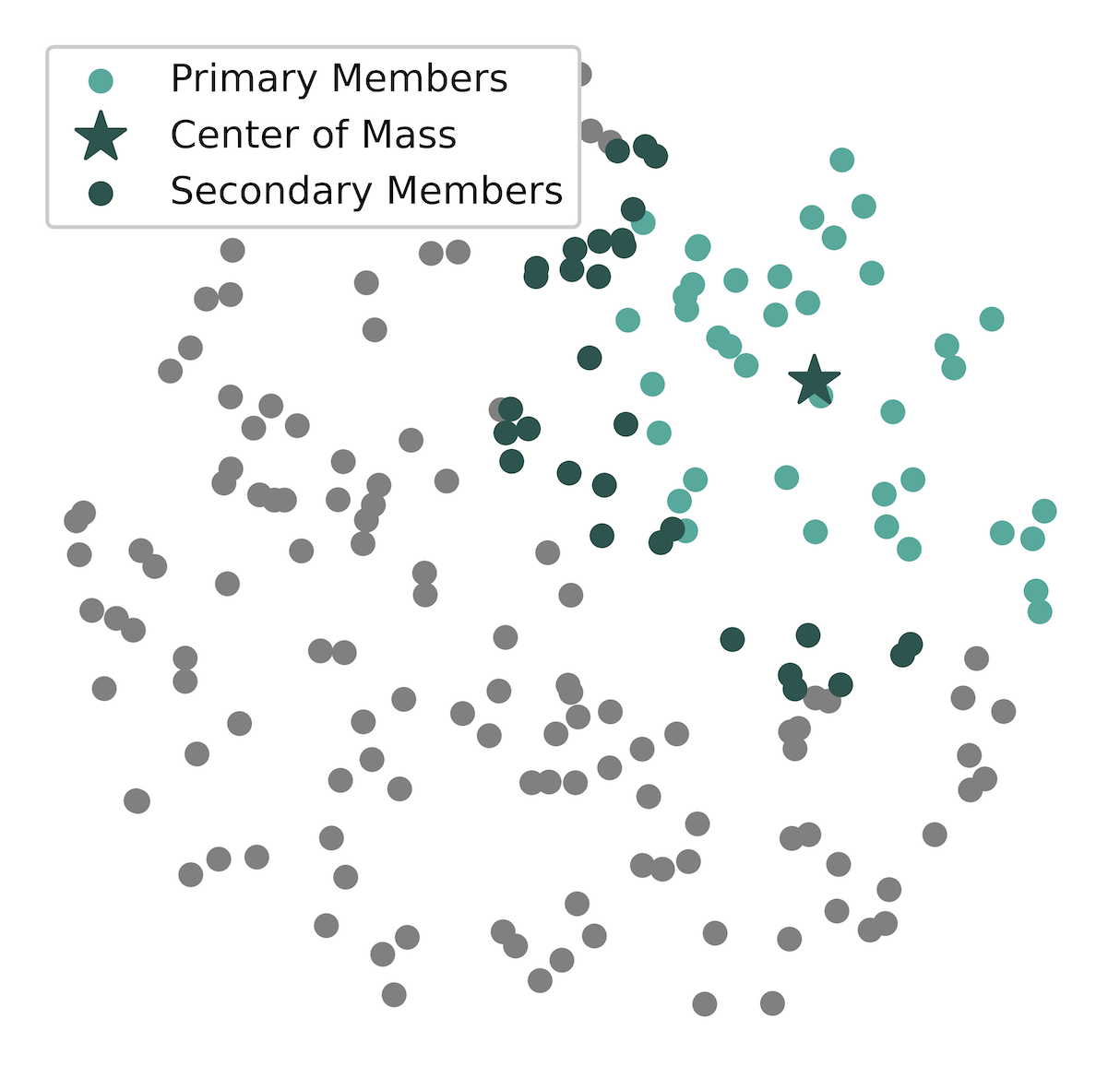}
    \caption{Example of assigning the community $C_3$ (Primary 2 in the legend) with $\hat{s}_3 = 40$ and $\eta = 1.75$ using a 2 dimensional reference layer. First, we select the element furthest from the origin that does not have a primary community as the seed. Next, we assign $C_3$ as the primary community of the seed, as well as the $\hat{s}_3 - 1$ nearest neighbours of the seed that do not yet have a primary community. Finally, we expand $C_3$ by a factor of $\eta$ (from $40$ to $40 \cdot 1.75 = 70$ nodes) by taking the nearest neighbours to the primary center of mass.}
    \label{fig:communities}
\end{figure}

At the end of the process, each element is assigned to exactly one primary community and some (possibly zero) secondary communities. In Figure~\ref{fig:communities} we show an example of the reference layer on $\hat{n}=150$ elements and three communities. Each of the three primary communities in this example consists of 50 elements before growing by a factor of $\eta=2.0$, attracting an additional 50 elements as its secondary members. From a computational perspective, to ensure that finding primary and secondary communities is efficient, we use k-d trees (short for k-dimensional tree) to perform spatial lookups of the sampled points~\cite{bentley1975kdtrees}.

\subsubsection*{Phase 4: assigning degrees to elements.}

We now have a degree sequence $(d_i, i \in [n])$, an assignment of degrees to outliers and non-outliers, and a collection of overlapping communities containing elements, each element belonging to one primary community and some number of secondary communities. Let $\hat{\textbf{d}}_{\hat{n}}$ be the subsequence (of length $\hat{n}$) of $(d_i, i \in [n])$ corresponding to the non-outliers. We are now ready to assign degrees in $\hat{\textbf{d}}_{\hat{n}}$ to the set of elements $R$ which will give us the non-outlier nodes of the graph. 

Similar to the potential problem with outliers, we want to avoid large degrees being assigned to small communities. When we create edges in Phase 5, element $v$, paired with some degree $d_i$, will have at least $\lfloor \xi d_i \rfloor$ of its degree delegated to the background graph, and will split the remainder of its degree evenly among its $\eta_v$ communities. Thus, if degree $d_i$ is paired with element $v$ then, for each community $C_j$ containing $v$, $|C_j| = s_j$ must be at least $\Big\lceil \lceil(1-\xi) d_i\rceil / \eta_v \Big\rceil + 1$ to accommodate the neighbours of $v$. This is in fact a lower bound as additional neighbours of $v$ coming from the background graph might end up in $C_j$ as well. This issue was addressed in the original \ABCD\ paper~\cite{kaminski2021artificial} where, to make room for the extra neighbours, a small correction was introduced, guided by the parameter 

\[
\phi = 1 - \sum_{k \in [L]} \left( \frac {\hat{s}_k}{\hat{n}} \right)^2 \frac { \hat{n} \xi }{ \hat{n} \xi + s_0 } \,.
\]
Typically, $\phi$ is very close to 1 and so this correction tends to be negligible in both theory and in practice. For consistency, we keep $\phi$ in the \ABCDoo\ model.

Before detailing the pairing process, let us explain how we deal with the correlation parameter $\rho$. Recall that our goal is to pair the $\hat{n}$ non-outlier degrees with the $\hat{n}$ elements such that the Pearson correlation between the degree and number of communities is approximately $\rho$. To this end, we perform multiple pairing attempts, with each attempt aiming to achieve a closer approximate the desired correlation. In each attempt, we fix a parameter $\alpha \in \R$ with negative values of $\alpha$ producing negative correlations and positive values producing positive correlations. Once the pairing is finished, we compute the obtained correlation $\rho_{\alpha}$. The parameter $\alpha$ is tuned to achieve correlation $\rho_{\alpha}$ close to the desired value $\rho$. This is achieved with a binary search, starting with $\alpha_{\min}=-60$ and $\alpha_{\max}=60$ and stopping once the desired precision is achieved or after the obtained correlation level stops improving. In some cases it is impossible to reach the correlation level required by the user, in which case the closest possible correlation is produced.

In each pairing attempt with parameter $\alpha \in \R$, we iteratively assign degrees of non-outliers to elements as follows. Recall that the non-outlier degree sequence $\hat{\textbf{d}}_{\hat{n}} = (\hat{d}_i, i \in [\hat{n}])$ is sorted with $\hat{d}_1$ being the maximum degree. Starting with $i=1$, let $U_i$ be the collection of unassigned elements at step~$i$. At step~$i$, choose element $v$ with probability proportional to $\eta_v^{\alpha}$ from the set of elements in $U_i$ satisfying 

\begin{equation}\label{eq:cond_degree}
\hat{d}_i \leq \frac{\eta_v}{1 - \xi \phi} \cdot \min \Big\{ |C_k|-1 : v \in C_k \Big\} \,,
\end{equation}
where recall that $\eta_v$ is the number of communities element $v$ belongs to. As a result, as desired, parameter $\alpha$ can be tuned to get negative correlation (negative~$\alpha$) as well as positive one (positive~$\alpha$) but the relationship between $\alpha$ and the correlation parameter $\rho$ is unclear. The chosen element $v$ gets paired with $\hat{d}_i$. If the set defined by the condition \eqref{eq:cond_degree} is empty, then we sample $v$ from the elements in $U_i$ for which the value of $({\eta_v}/({1-\xi \phi})) \cdot \min \{ |C_k|-1 : v \in C_k \}$ is maximal. 

Once the pairing is complete, we relabel as follows. The nodes are labelled as $[n]$, with node $i$ having degree $d_i$ (note that the inequality $d_1 \geq \dots \geq d_n$ still holds). The communities $(C_j, j \in L)$ now contain non-outlier nodes corresponding to degree/element pairing. If $d_i$ was assigned a non-outlier, its corresponding node $i$ belongs to some number $\eta_i$ of communities, one of which is its primary community. If $d_i$ was assigned to an outlier, its corresponding node $i$ is assigned to an auxiliary ``community'' labelled $C_0$; the set $C_0$ is merely a label for the outliers and such an outlier $i$ has $\eta_i = 0$.

\subsubsection*{Phase 5: creating edges.}

At this point there are $n$ nodes with labels from $[n]$; $\hat{n}=n-s_0$ of them are non-outliers and the remaining ones are outliers. There is also a family of overlapping communities with each non-outlier node $i \in [n]$ belonging to $\eta_i \ge 1$ communities. Finally, each node $i \in [n]$ (either outlier or non-outlier) is assigned a degree $d_i$ which we interpret as a set of $d_i$ unpaired half-edges. The remaining 2 steps construct the edges. 

For each non-outlier $i \in [n]$ we split the $d_i$ half-edges of $i$ into \textit{community} half-edges and \textit{background} half-edges. To this end, define $Y_i := \round{(1-\xi) d_i}$ and $Z_i := d_i - Y_i$ (note that $Y_i$ and $Z_i$ are random variables with $\E{Y_i} = (1-\xi) d_i$ and $\E{Z_i} = \xi d_i$) and, for all non-outliers $i \in [n]$, split the $d_i$ half-edges of $i$ into $Y_i$ community half-edges and $Z_i$ background half-edges. Community half-edges are further split into the $\eta_i$ communities the non-outlier node $i$ belongs to, as evenly as possible. Specifically, for the communities containing node $i$, $Y_i - \eta_i \lfloor Y_i/\eta_i \rfloor$ communities (chosen randomly) each receive $\lfloor Y_i/\eta_i \rfloor + 1$ half-edges and the remaining communities each receive $\lfloor Y_i/\eta_i \rfloor$ half-edges. Note that condition~(\ref{eq:cond_degree}) makes sure that after splitting the community half-edges into communities, as evenly as possible, there is enough room in \emph{each} community. On the other hand, if $i \in [n]$ is an outlier then we set $Z_i = d_i$.

Once the assignment of degrees is complete, for each $j \in [L]$, we independently construct the \textit{community graph} $G_{n,j}$ as follows. For each node $i \in C_j$, there is a set of half-edges attached to $i$ delegated to $C_j$. With these delegated half-edges, we form $G_{n, j}$ as per the configuration model (see Subsection~\ref{subsec:config} for the definition of the model and relevant references). In the event that the sum of degrees in a community is odd, we pick a maximum degree node $i$ in said community and decrease its community degree by one while increasing its background graph degree by one. Finally, construct the \textit{background graph} $G_{n,0}$ as per the configuration model on node set $[n]$ and degree sequence $(Z_i,i \in [n])$. Let $G_n = \bigcup_{0 \leq j \leq n} G_{n,j}$ be the union of all graphs generated in this phase.

\subsubsection*{Phase 6: rewiring self-loops and multi-edges.}

Note that, although we are calling $G_{n,0},G_{n,1},\dots,G_{n,L}$ \textit{graphs}, they are in fact \textit{multi-graphs} at the end of phase~5. To ensure that $G_n$ is simple, we perform a series of \textit{rewirings} in $G_n$. A rewiring takes two edges as input, splits them into four half-edges, and creates two new edges distinct from the input. We first rewire each community graph $G_{n,j}$ ($j \in [L]$), and the background graph $G_{n, 0}$, independently as follows.
\begin{enumerate}
\item For each edge $e \in E(G_{n,j})$ that is a loop, we add $e$ to a \textit{recycle} list that is assigned to $G_{n,j}$. Similarly, if $e\in E(G_{n,j})$ contributes to a multi-edge, we put all but one copies of this edge to the \textit{recycle} list.
\item We shuffle the \textit{recycle} list and, for each edge $e$ in the list, we choose another edge $e'$ uniformly from $E(G_{n,j}) \setminus \{e\}$ (not necessarily in the list) and attempt to rewire these two edges. We save the result only if the rewiring does not lead to any further self-loops or multi-edges, otherwise we give up. In either case, we then move to the next edge in the \textit{recycle} list. 
\item After we attempt to rewire every edge in the \textit{recycle} list, we check to see if the new \textit{recycle} list is smaller. If yes, we repeat step 2 with the new list. If no, we give up and move all of the ``bad'' edges from the community graph to a collective \textit{global recycle} list.
\end{enumerate}

As a result, after ignoring edges in the \textit{global recycle} list, all community graphs are simple and the background graph is simple. However, as is the case in the original \ABCD\ model, an edge in the background graph can form a multi-edge with an edge in a community graph. Another problem that might occur, specific to \ABCDoo\ model, is that an edge from one community can form a multi-edge with an edge from a different but overlapping community. All of these problematic edges are added to the \emph{global recycle} list. We merge all community graphs with the background graph. Finally, the \textit{global recycle} list is transformed into a list of half-edges and new edges are created from it. We follow the same procedure as for the community graphs. As the background graph is sparse, this final rewiring is very fast in practice. However, in rare cases it is possible that the process does not terminate. We try rewiring 100 times the size of the global recycle list. If, after this many iterations, we cannot find any new admissible edge (thus reducing the size of global recycle list) we throw an error and indicate to the user that the algorithm failed to terminate with success for the given input parameters (indicating that the likely cause is that the required input degree sequence was not graphic and it should be reconsidered by the user).

%%%%%%%%%%%%%%%%%%%%%%%%%%%%%%%%%%%%%%%%%%%%%%%%%%%%%%%%%%%
\section{Properties of the \ABCDoo\ Model}\label{sec:properties of model}
%%%%%%%%%%%%%%%%%%%%%%%%%%%%%%%%%%%%%%%%%%%%%%%%%%%%%%%%%%%

In this section, we present experiments highlighting the properties of the \ABCDoo\ model. We begin by investigating the communities generated by the reference layer as described in Section~\ref{sec:effects_of_reference_layer}. We then continue with additional features of the model in Section~\ref{sec:graph_properties}. We are mainly interested in comparing the model's properties with properties of real-world networks. Thus, for the model parameters, we use values derived from several networks that have known overlapping communities. The networks were presented in~\cite{yang2012defining}, and are available on the Stanford Network Analysis Project (SNAP) website\footnote{\url{https://snap.stanford.edu/data/index.html\#communities}}.

\bigskip
\noindent
\textbf{DBLP:} a network where nodes are authors and an edge exists between two authors if they published a paper together. The ground-truth communities are defined by the publication venues of the journals in which the authors published.

\smallskip
\noindent
\textbf{Amazon:} a network where nodes are products and an edge exists between two products if they are frequently co-purchased.
The ground-truth communities are defined by product categories. 

\smallskip
\noindent
\textbf{YouTube:} a network where nodes are YouTube channels and an edge exists between two channels if they are friends on the platform. The ground-truth communities are user-generated social groups. In the full network over $95\%$ of nodes are outliers and the level of noise $\xi$ is $0.96$. We consider the subgraph induced by the non-outlier nodes to avoid such extreme values.

\bigskip
For each network, all parameters other than $\delta$ and $s$ are measured directly from the data.
We set $\delta = 5$ to avoid cases where nodes have no neighbours in their own community, and insist on $s \geq 10$. The power-law exponents $\gamma$ and $\beta$ are fit using maximum-likelihood estimation described in \cite{clauset2009powerlaw} and implemented in the Python package \textbf{powerlaw}\footnote{\url{https://github.com/jeffalstott/powerlaw}}\cite{alstott2014powerlaw}.
In this fitting process, we set the minimum degree as $\delta = 5$ and the minimum community size as $s=10$ for consistency. The level of noise $\xi$ is computed directly from the graphs as the fraction $|E_b|/|E|$ where $E_b$ is the set of edges where the incident nodes share no communities. The parameters for each network are shown in Table~\ref{tab:real_graph_params2}.
Additionally, most of the experiments are affected by the dimension $d$ of the hidden reference layer. In such cases, we compare dimensions $2$, $8$, and $64$. 

\begin{table}[ht]
    \centering
    \begin{tabular}{|c|p{0.52\linewidth}|c|c|c|}
        \hline
         Parameter & Description & DBLP & Amazon & YouTube \\ \hline
         $n$ & Number of nodes & $317{,}080$ & $334{,}863$ & $52{,}675$ \\
         $s_0$ & Number of outliers & $56{,}082$ & $17{,}669$ & $0$ \\
         $\eta$ & Average number of communities a non-outlier node is part of & $2.76$ & $7.16$ & $2.45$ \\
         $\rho$ & Pearson correlation between the degree of nodes and the number of communities they belong to & $0.76$ & $0.22$ & $0.37$ \\ \hline
         $\gamma$ & Power-law degree distribution with exponent $\gamma$ & $2.30$ & $3.04$ & $1.87$ \\
         $\delta$ & Min degree at least $\delta$ & $5$ & $5$ & $5$ \\
         $\Delta$ & Max degree at most $\Delta$ & $343$ & $549$ & $1{,}928$\\
         \hline
         $\beta$ & Power-law community size distribution with exponent $\gamma$ & $1.88$ & $2.03$ & $2.13$\\
         $s$ & Min community size at least $s$ & $10$ & $10$ & $10$\\
         $S$ & Max community size at most $S$ & $7{,}556$ & $53{,}551$ & $3{,}001$ \\
         \hline
         $\xi$ & Level of noise & $0.11$ & $0.11$ & $0.59$ \\ \hline
    \end{tabular}
    \medskip
    \caption{The parameters used for generating \ABCDoo{} graphs based on the DBLP, Amazon, and YouTube datasets.}
    \label{tab:real_graph_params2}
\end{table}

\subsection{Effects of the Reference Layer} \label{sec:effects_of_reference_layer}

The new and novel feature of the \ABCDoo\ model is the hidden reference layer used to generate overlapping communities. Here, we analyze the effects of this reference layer on the community overlap sizes. For comparison, we consider the \textbf{CKB} model~\cite{chykhradze2014ckb} following a description from~\cite{sengupta2017ckbdescription}. The \textbf{CKB} model is an application of the random bipartite community affiliation graph proposed for overlapping \textbf{LFR}~\cite{lancichinetti2009benchmarks} that uses a power-law distribution for the number of communities per node. Let us briefly summarize this graph construction process. The community affiliation graph is a bipartite graph where one part consists of nodes and the other consists of communities. As input, the \textbf{CKB} model requires the number of nodes $n$, a truncated power-law distribution $\tpl{\Omega}{x_{\min}}{x_{\max}}$ for the number of communities per node, and another truncated power-law distribution $\tpl{\beta}{s}{S}$ for the community sizes. The number of communities is set to 
\[
\left\lfloor \frac{n \cdot \e[\tpl{\Omega}{x_{\min}}{x_{\max}}]}{\e[\tpl{\beta}{s}{S}]} \right\rfloor \,.
\]

First, each node in the bipartite graph is assigned a number of half-edges sampled from $\tpl{\Omega}{x_{\min}}{x_{\max}}$, and likewise each community is assigned a number of half-edges sampled from $\tpl{\beta}{s}{S}$. A small adjustment is made to ensure that the number of half-edges in both parts is equal. Lastly, the nodes are matched with the communities based on the bipartite configuration model. Thus, each community is populated with nodes, and each node is assigned to some number of communities. 

\subsubsection*{Community sizes.}

\begin{figure}[ht]
    \centering
    \includegraphics[width=0.48\linewidth]{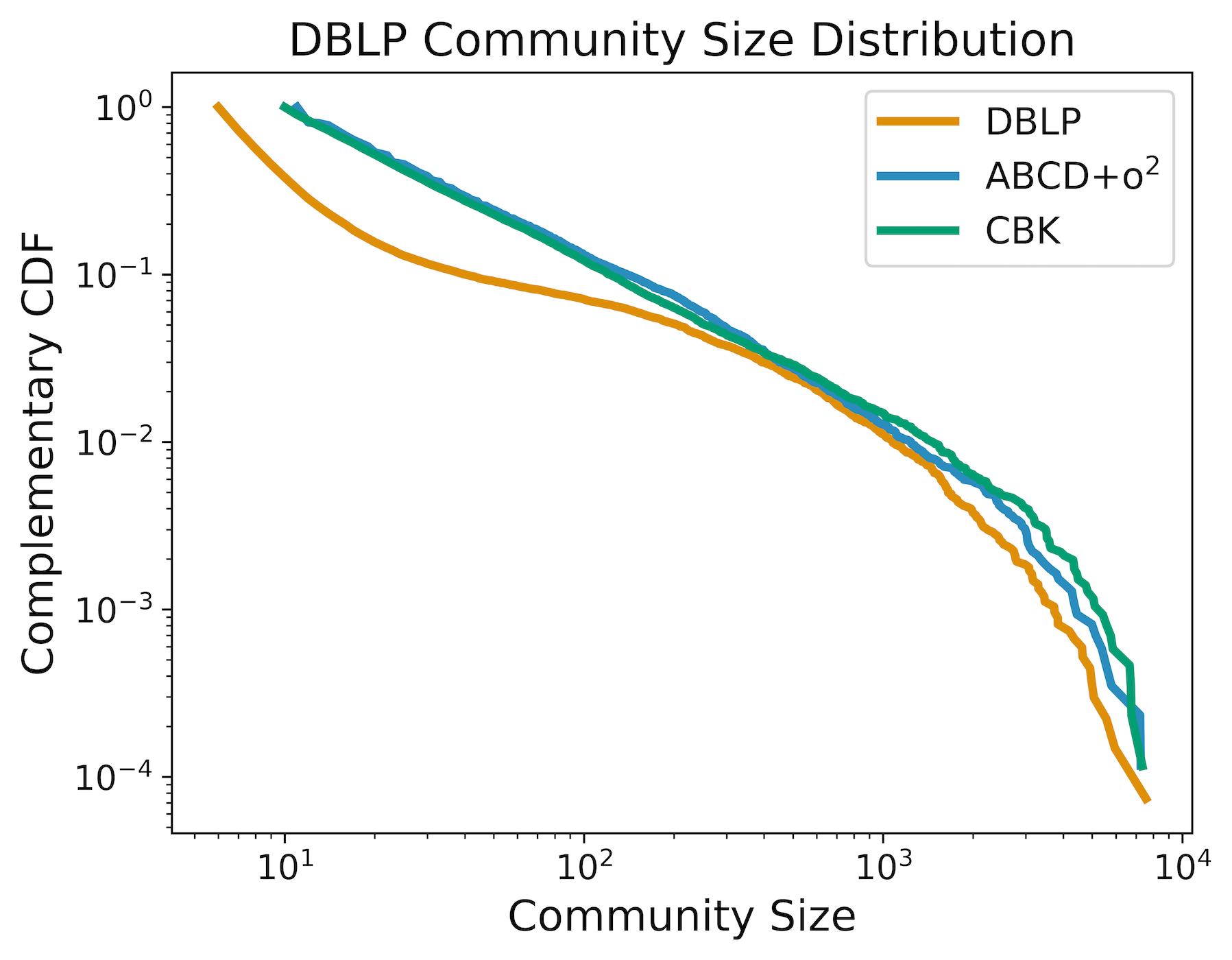}\hfill
    \includegraphics[width=0.48\linewidth]{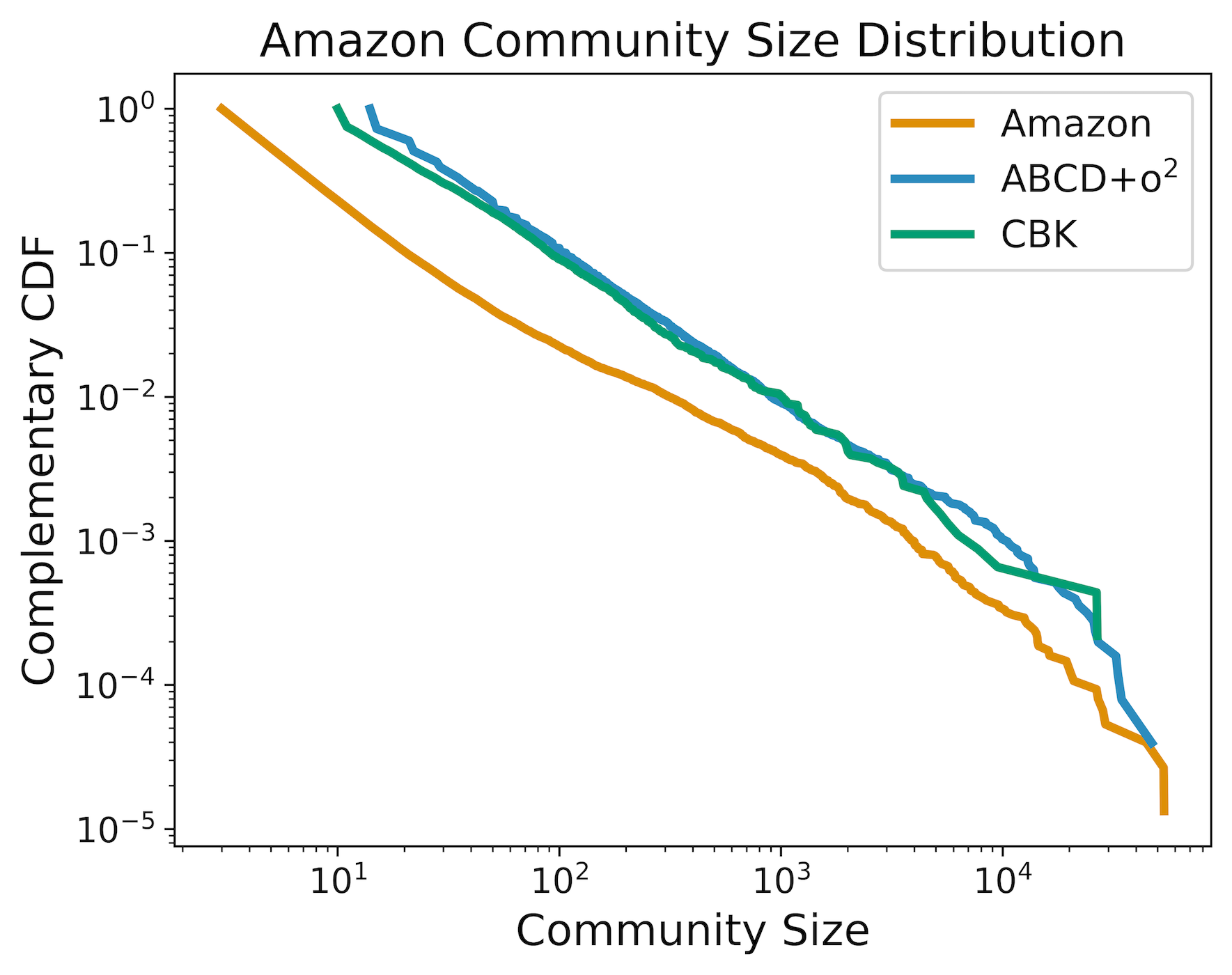}\\
    \includegraphics[width=0.48\linewidth]{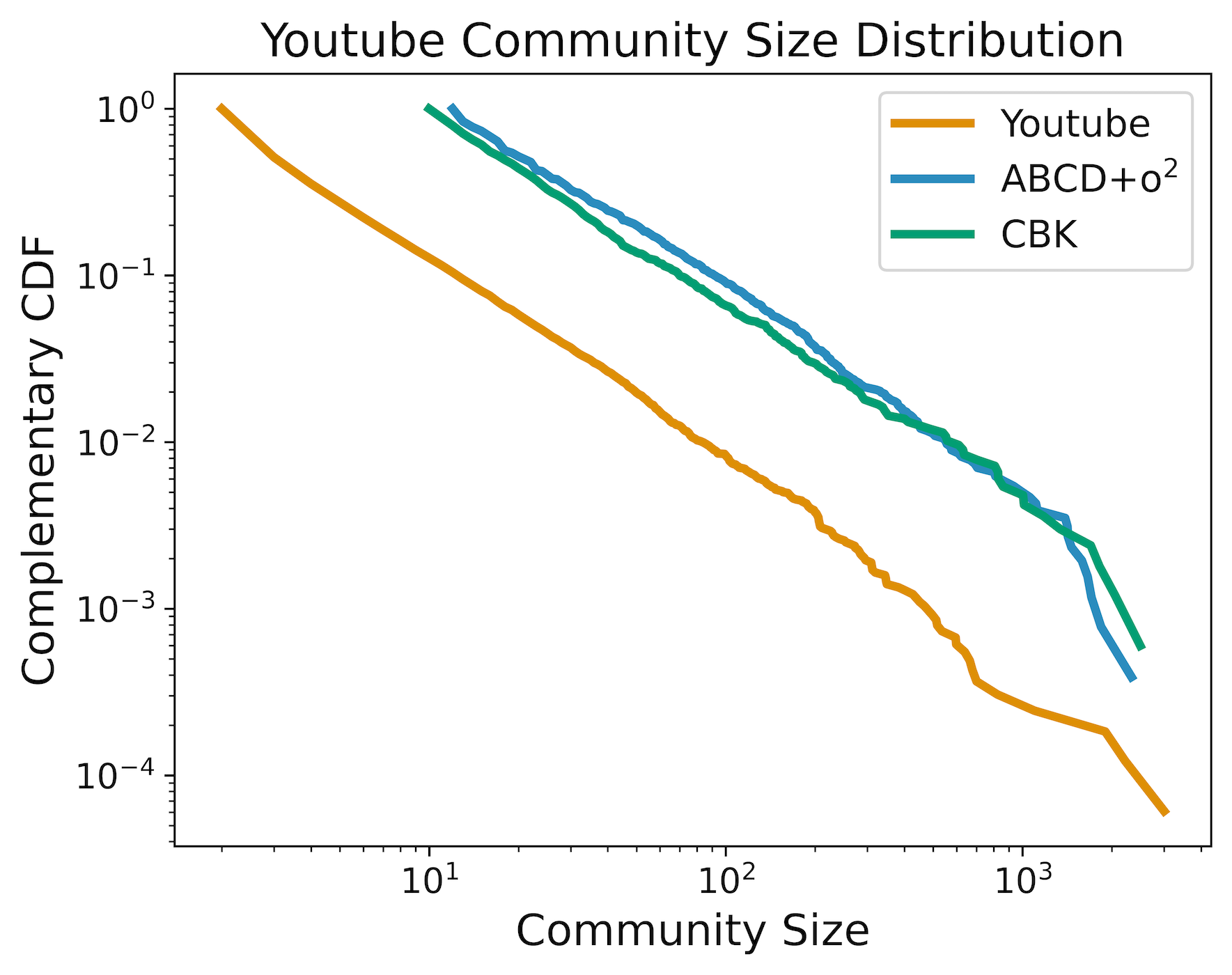}
    \caption{Distribution of community sizes for the three real-world networks and their synthetic counterparts. The y-axis shows the empirical complementary cumulative distribution function, computed as $y = 1 - |\{C_i : |C_i| < x\}| \; / \; |\{C_i\}|$.}
    \label{fig:com-size}
\end{figure}

In the first experiment (see Figure~\ref{fig:com-size}), we compare the distributions of community sizes. Of course, since both \ABCDoo\ and \textbf{CKB} sample community sizes directly from the specified distribution, it is unsurprising that the empirical distributions are near-perfect fits when accounting for the imposed minimum community size of $10$. Thus, this first experiment merely acts as a sanity check ensuring that the \ABCDoo\ and \textbf{CKB} models generate community sizes correctly. Moreover, this experiment verifies that the distributions coming from the real networks are indeed power-law.

\subsubsection*{Number of communities per node.}
\begin{figure}[ht]
    \centering
    \includegraphics[width=0.48\linewidth]{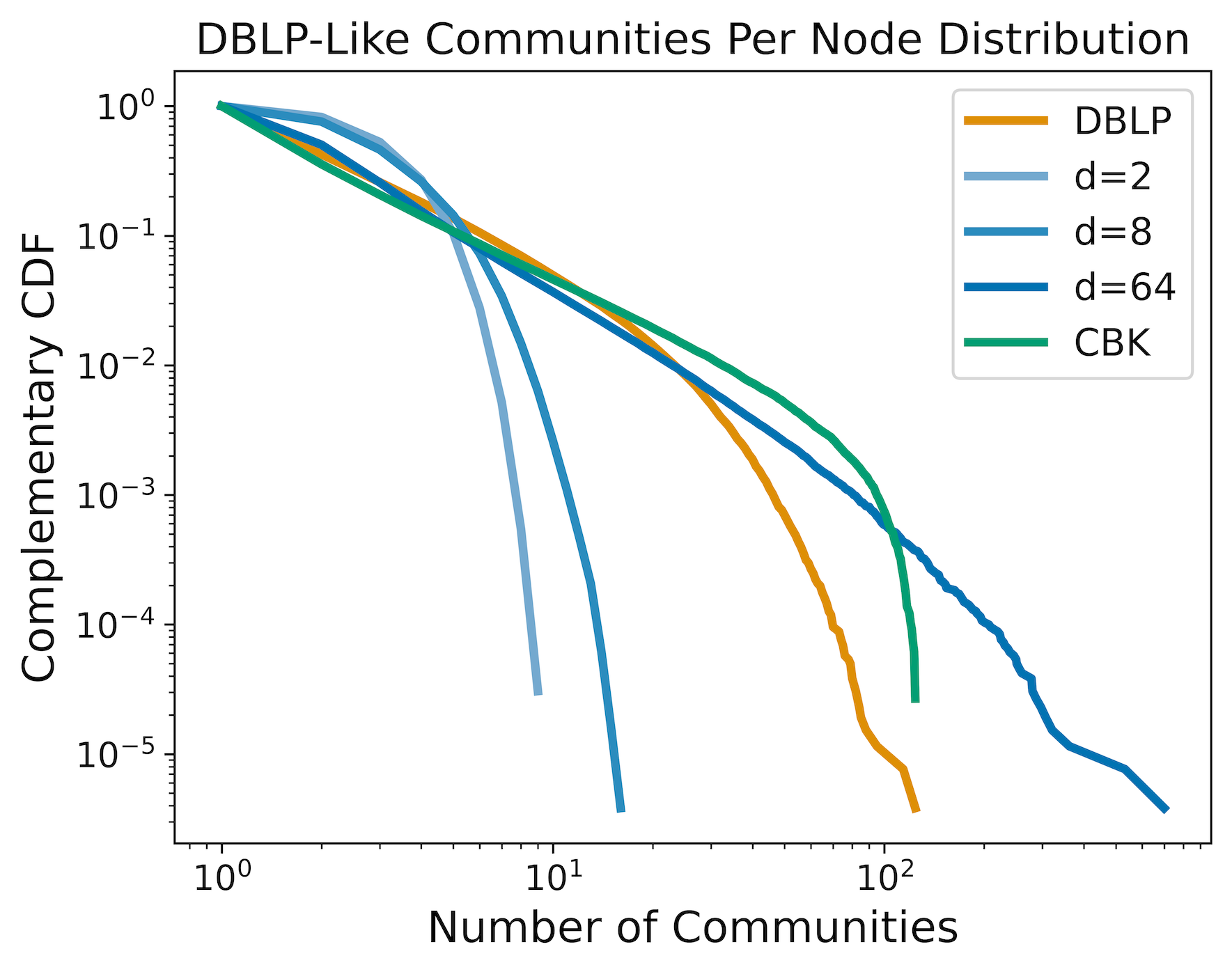}\hfill
    \includegraphics[width=0.48\linewidth]{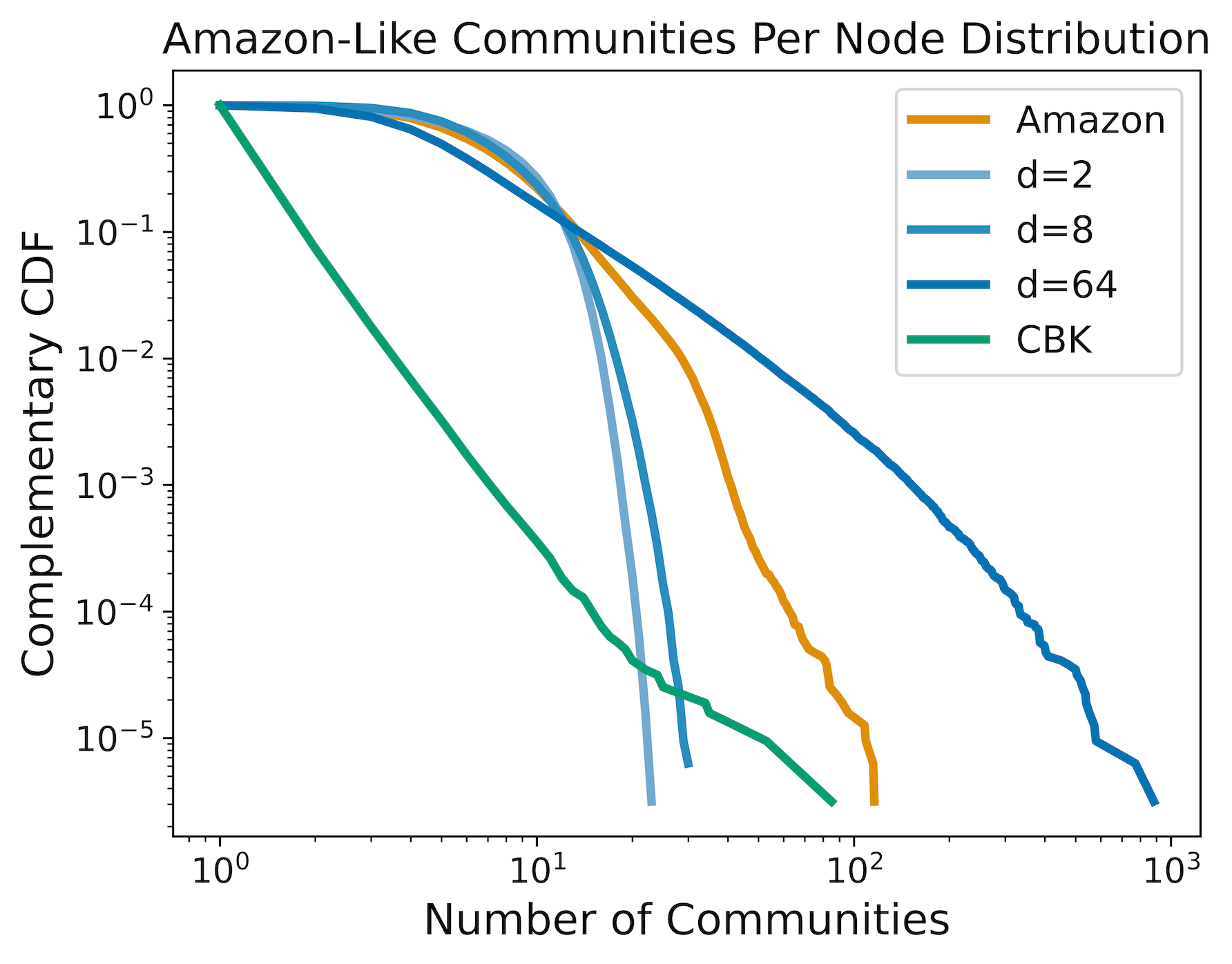}\\
    \includegraphics[width=0.48\linewidth]{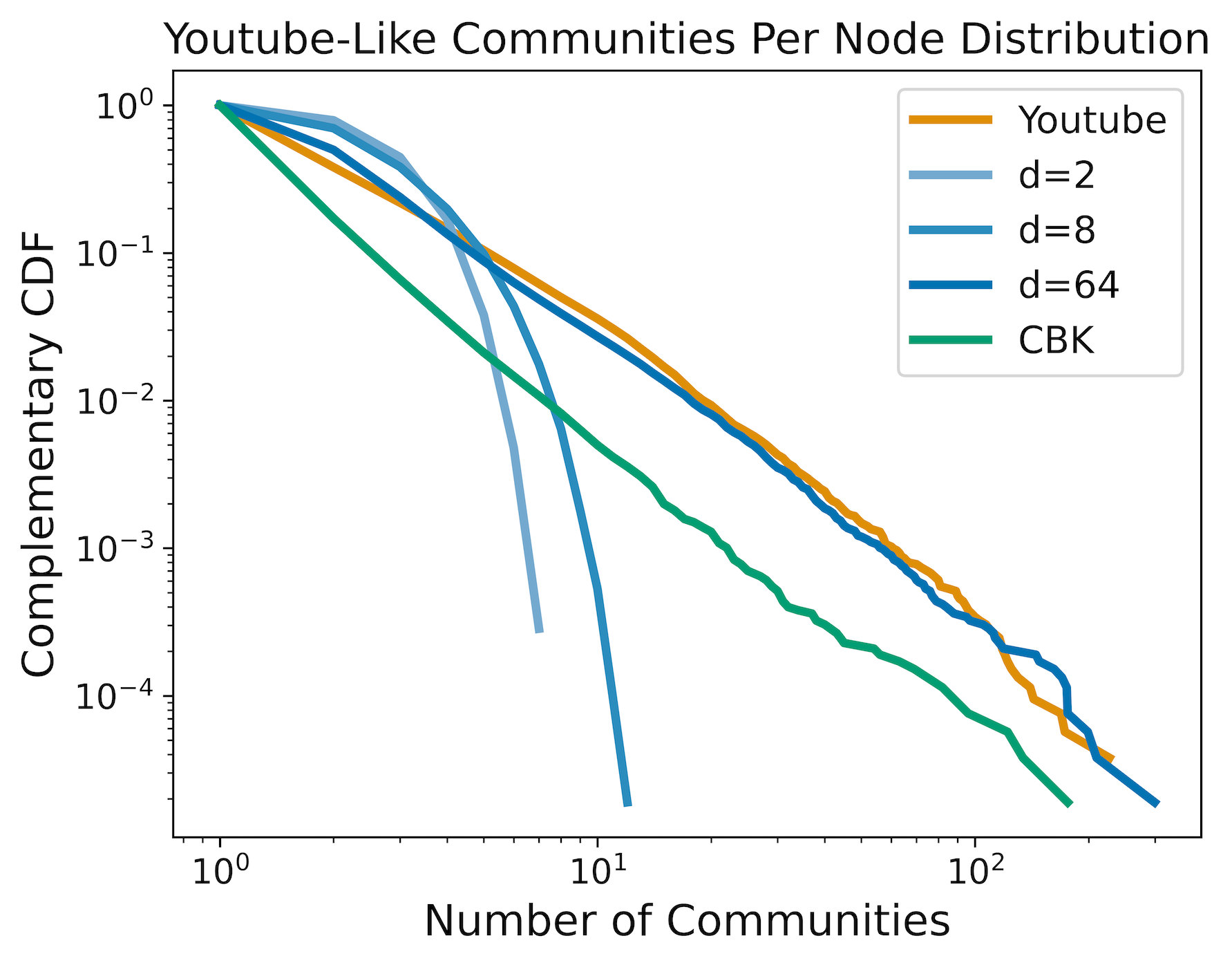}
    \caption{Distribution of the number of communities per node. The y-axis shows the empirical complementary cumulative distribution function, computed as $y = 1 - |\{v_i : |\{C_j : v_i \in C_j\} < x\}| \; / \; |V|$.}
    \label{fig:coms-per-node}
\end{figure}

Results are presented in Figure~\ref{fig:coms-per-node}. In contrast with the previous experiment, there is less agreement between the two models and the real distributions. In the case of the YouTube and DBLP networks, the power-law distribution generated by the \textbf{CKB} model is a good fit for the empirical distribution, whereas the Amazon distribution does not appear to be power-law. In contrast, the \ABCDoo\ model is able to produce a distribution similar to each of the three networks with a properly tuned dimension parameter $d$; for the Amazon graph a low dimensional reference layer captures the behaviour well, and for the DBLP and YouTube graphs a high dimensional reference layer captures the behaviour well. The flexibility and realistic distributions are even more encouraging when we recall that the distribution is not given to the \ABCDoo model, and is instead a natural feature of the hidden reference layer. 

\subsubsection*{Community intersection sizes.}
\begin{figure}[ht]
    \centering
    \includegraphics[width=\linewidth]{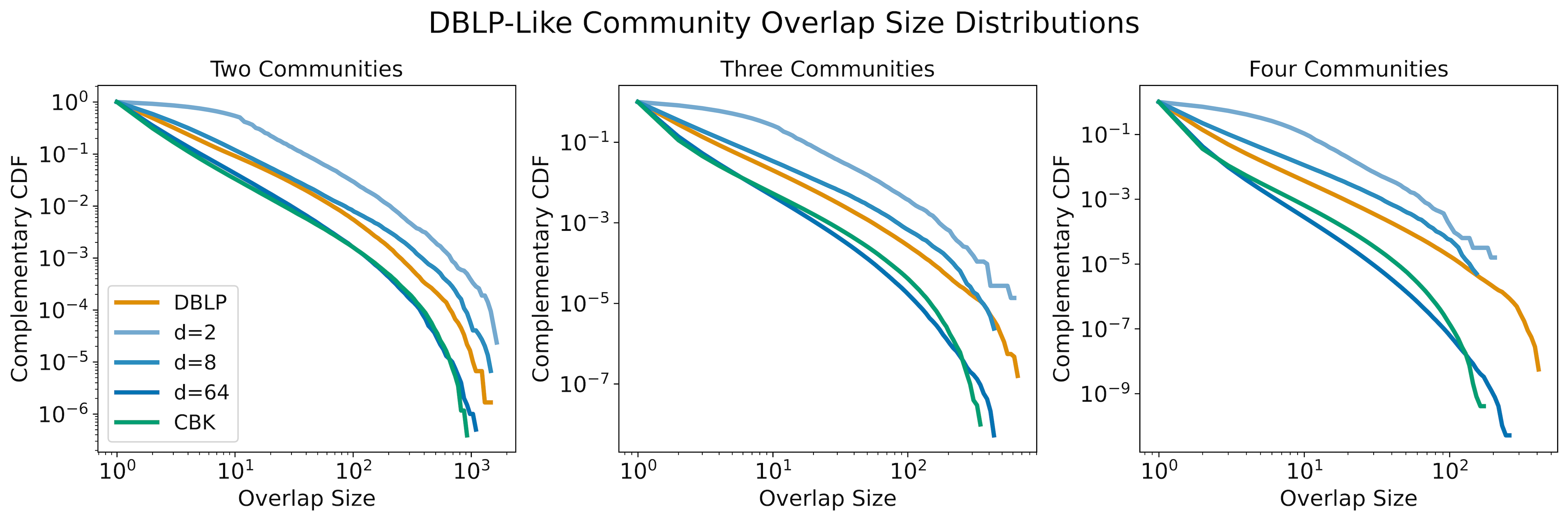} \\
    \includegraphics[width=\linewidth]{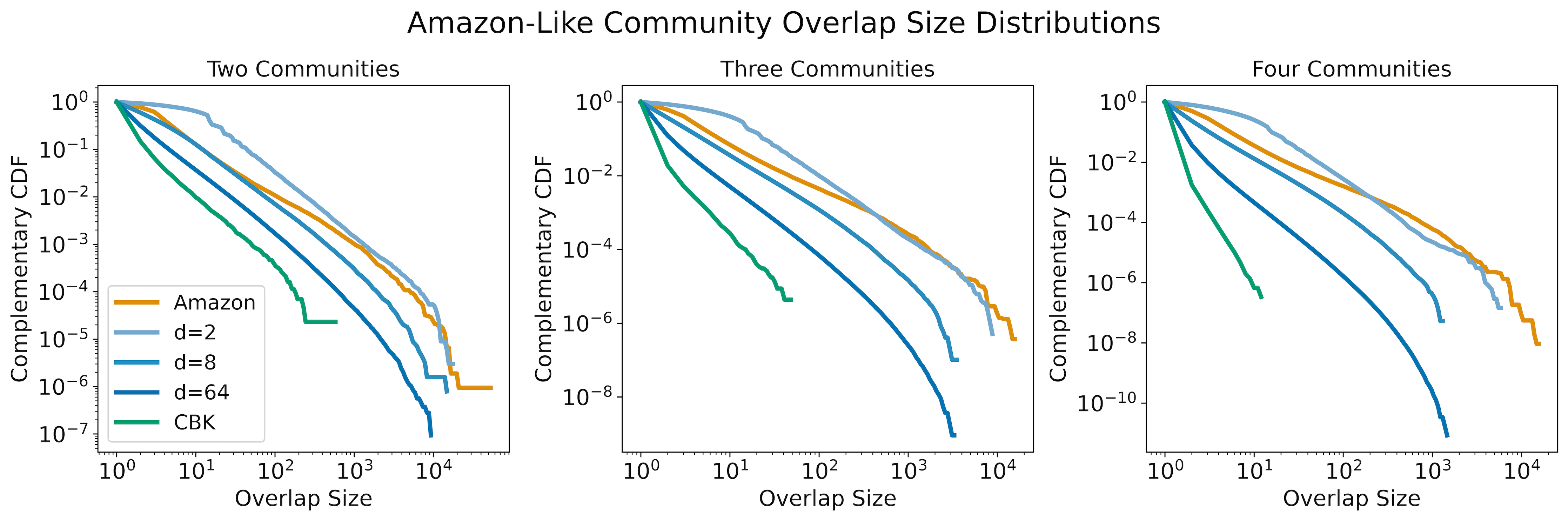} \\
    \includegraphics[width=\linewidth]{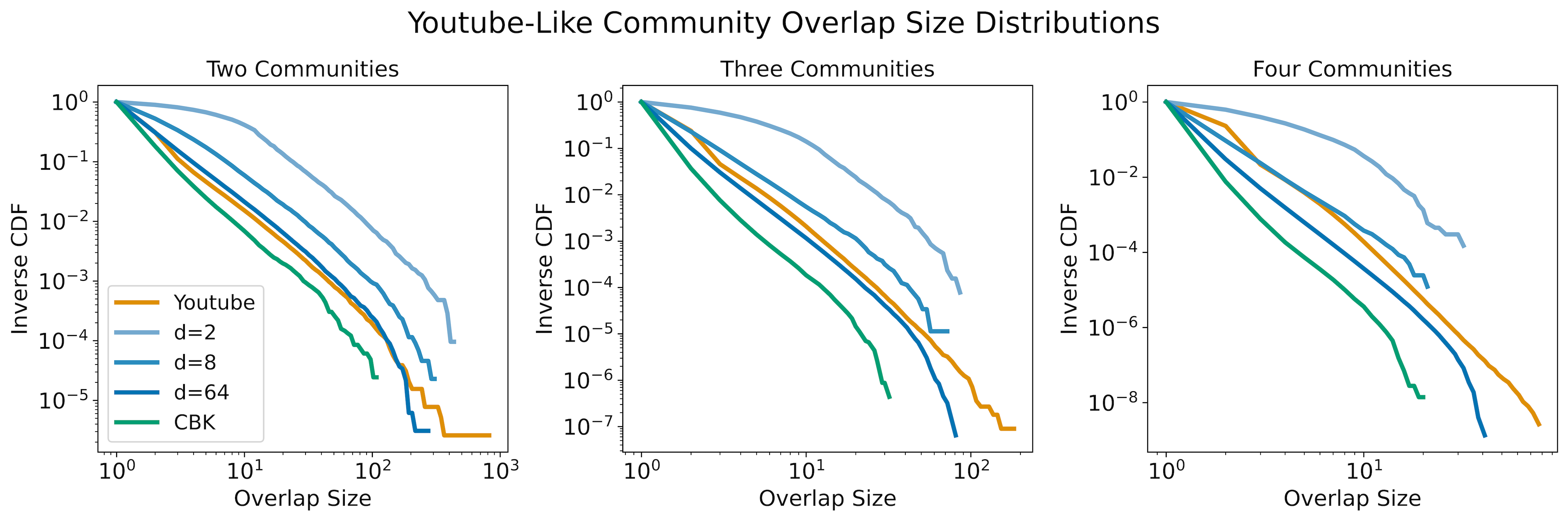}
    \caption{Distribution of the size of overlaps. The y-axis shows the empirical complementary cumulative distribution function of non-empty overlaps. For example, the center column of figures corresponds to three overlapping communities and is computed as $y = 1 - |\{ \{C_i, C_j, C_k\} : |C_i \cap C_j \cap C_k| < x\}| \; / \; |\{ \{C_i, C_j, C_k\} : |C_i \cap C_j \cap C_k| \neq \emptyset|$.}
    \label{fig:com-overlap}
\end{figure}

Results are presented in Figure~\ref{fig:com-overlap}. We examine the size of the overlaps produced by each model. Neither the \ABCDoo\ model nor the \textbf{CKB} model specifies a particular distribution, but we can see the \ABCDoo{} model, with a well-tuned dimension parameter, is able to create a similar distribution to the real graph. In contrast, the \textbf{CKB} model does not fit the Amazon and YouTube graphs well. Specifically, the \textbf{CKB} model produces too many small intersections and, moreover, the largest intersection is several orders of magnitude too small. We consider this experiment strong evidence for using a reference layer as the default option for generating overlapping communities.

\subsection{Additional properties of the model.} \label{sec:graph_properties}

We now turn to properties of the \ABCDoo\ model beyond those governed by the hidden reference layer.

\subsubsection*{Degree vs.\ number of communities.}

The \ABCDoo\ model attempts to produce a correlation $\rho$ between the degree of a node and the number of communities it belongs to. In Table~\ref{tab:rho} we report the desired correlation $\rho$ for each of the networks, as well as the achieved correlation for the \ABCDoo\ graphs. Note that the extreme positive correlation found in the DBLP network cannot be matched by the model, although it can still achieve a strong positive correlation. The maximum achievable correlation increases with the dimension as there are more nodes belonging to a large number of communities (see Figure~\ref{fig:coms-per-node}).

\begin{table}[ht]
    \centering
    \begin{tabular}{|c|c|c|c|c|}
        \hline
        Graph & Empirical & $d=2$ & $d=8$ & $d=64$ \\ \hline
        DBLP & $0.76$ & $0.42$ & $0.56$ & $0.68$ \\
        Amazon & $0.22$ & $0.20$ & $0.19$ & $0.20$ \\
        YouTube & $0.37$ & $0.37$ & $0.36$ & $0.38$ \\
        \hline
    \end{tabular}
    \smallskip
    \caption{Measured $\rho$ in both the real networks and the corresponding \ABCDoo\ graphs. We generated 10 \ABCDoo\ graphs for each dimension, and found that across all graphs and dimensions, the standard error, to two decimal places, was $0$.}
    \label{tab:rho}
\end{table}

\subsubsection*{Density of intersections}
\begin{figure}[ht]
    \centering
    \includegraphics[width=0.48\linewidth]{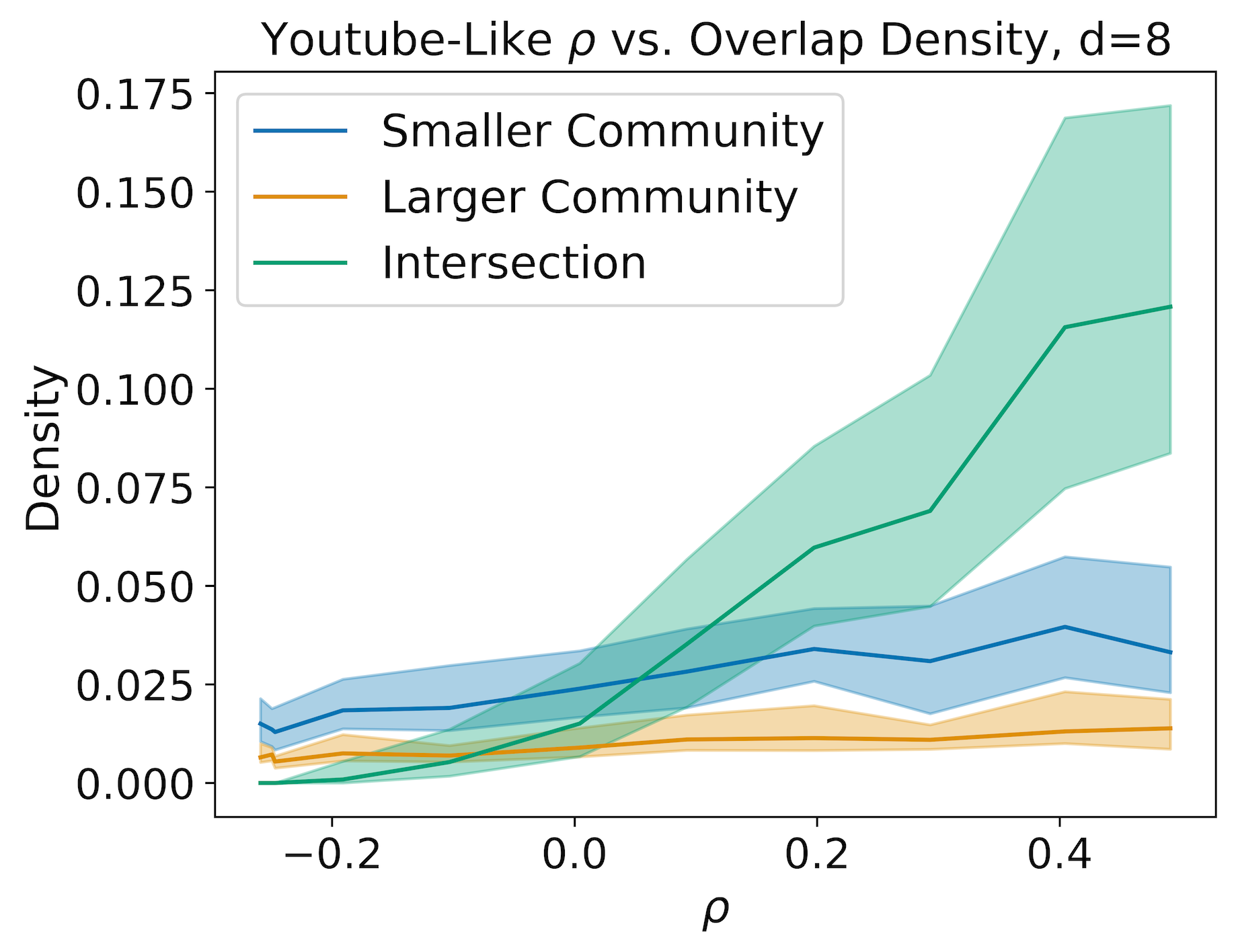}
     \includegraphics[width=0.48\linewidth]{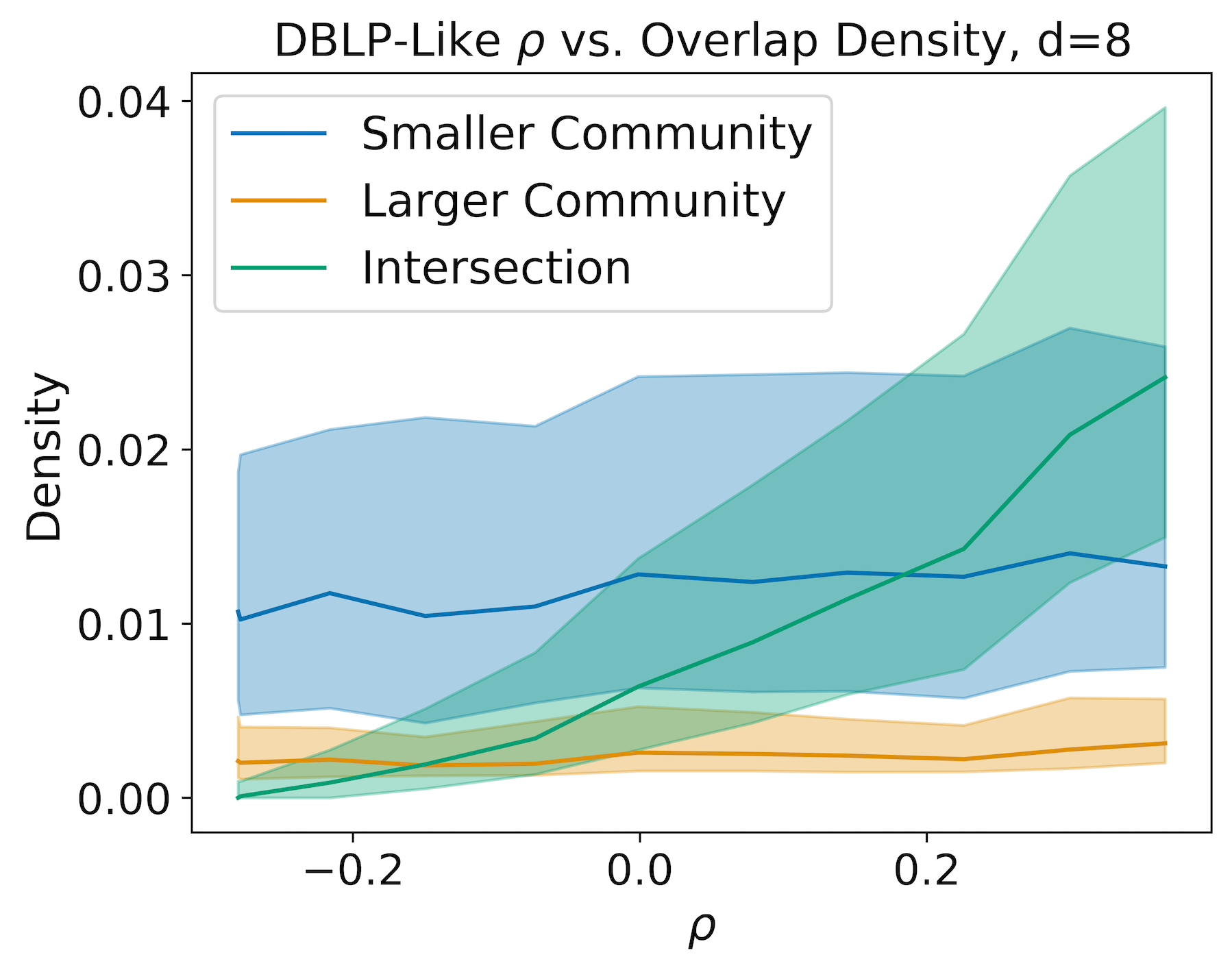}
    \caption{Community overlap densities compared to individual community densities for various values of $\rho$. For 9 \ABCDoo\ graphs with $\rho$ evenly spaced in $[-0.5, 0.5]$, we compute the empirical value $\hat{\rho}$ and the distribution of density for overlapping communities. We plot $\hat{\rho}$ against the distribution of the density for the intersection and each community individually. The line corresponds to the median, and the shaded region to the 25th to 75th quantile. For computational reasons, we only consider overlaps of size at least 25, and to ensure we are not simply measuring the density of the smaller community, the size of the overlap must be at most half the size of the smaller community.}
    \label{fig:intersection density vs rho}
\end{figure}

One consequence of forcing nodes of large degree into community intersections is that those intersections will be denser than the individual communities. We show this phenomenon in Figure~\ref{fig:intersection density vs rho} by comparing the community and intersection densities for various $\rho$ while fixing the other parameters. Note that, given communities $C_j$ and $C_k$ with $s_j \gg s_k$, the density of $C_k$ is almost always larger than that of $C_j$ since both community graphs are sparse (at most a $1-\xi$ fraction of each node's degree is delegated to one of its communities). Figures for the other graphs and dimensions can be found in Appendix~\ref{a:intersection_density}.

\begin{figure}[ht]
    \centering
    \includegraphics[width=0.48\linewidth]{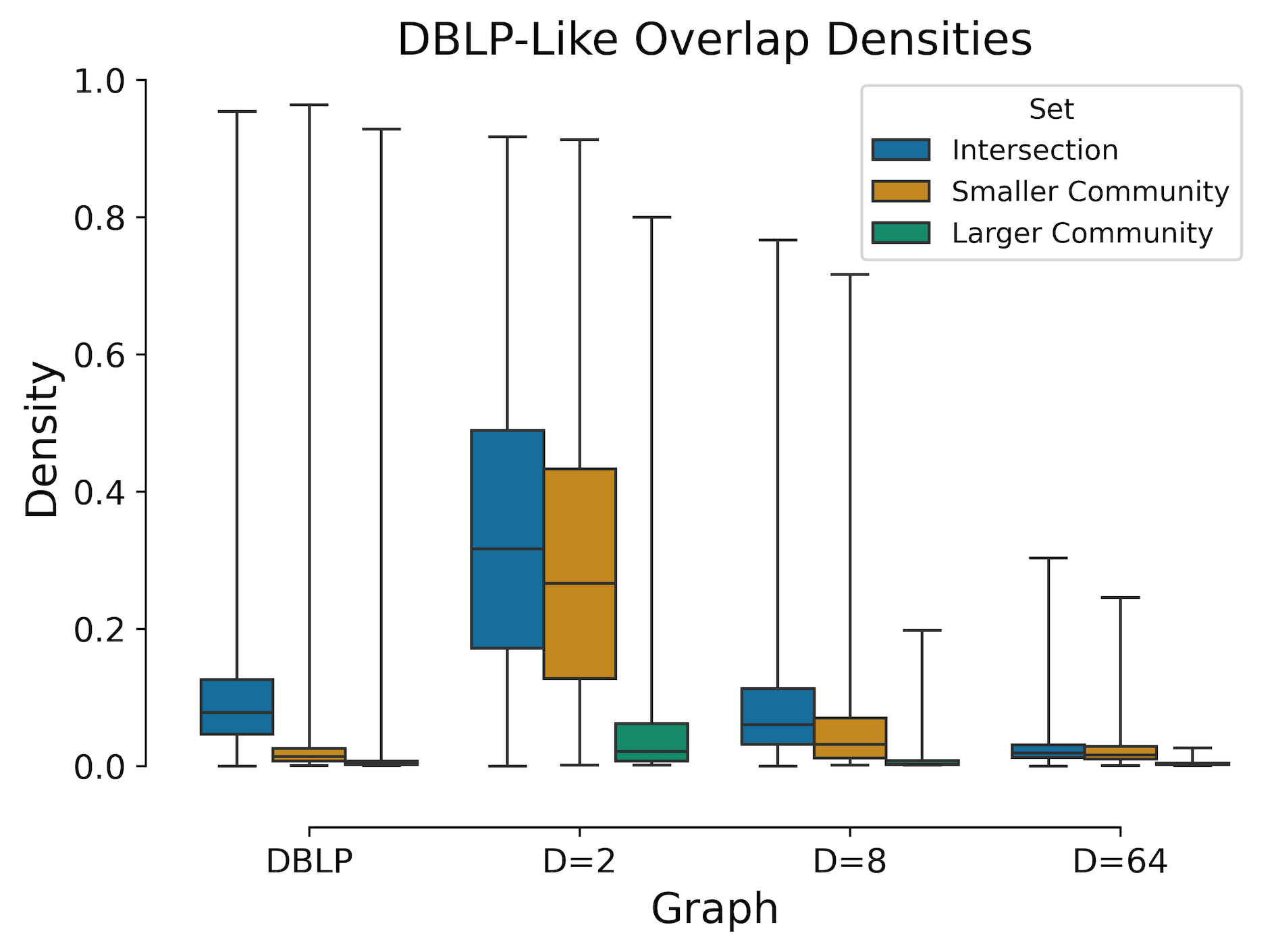}\hfill
    \includegraphics[width=0.48\linewidth]{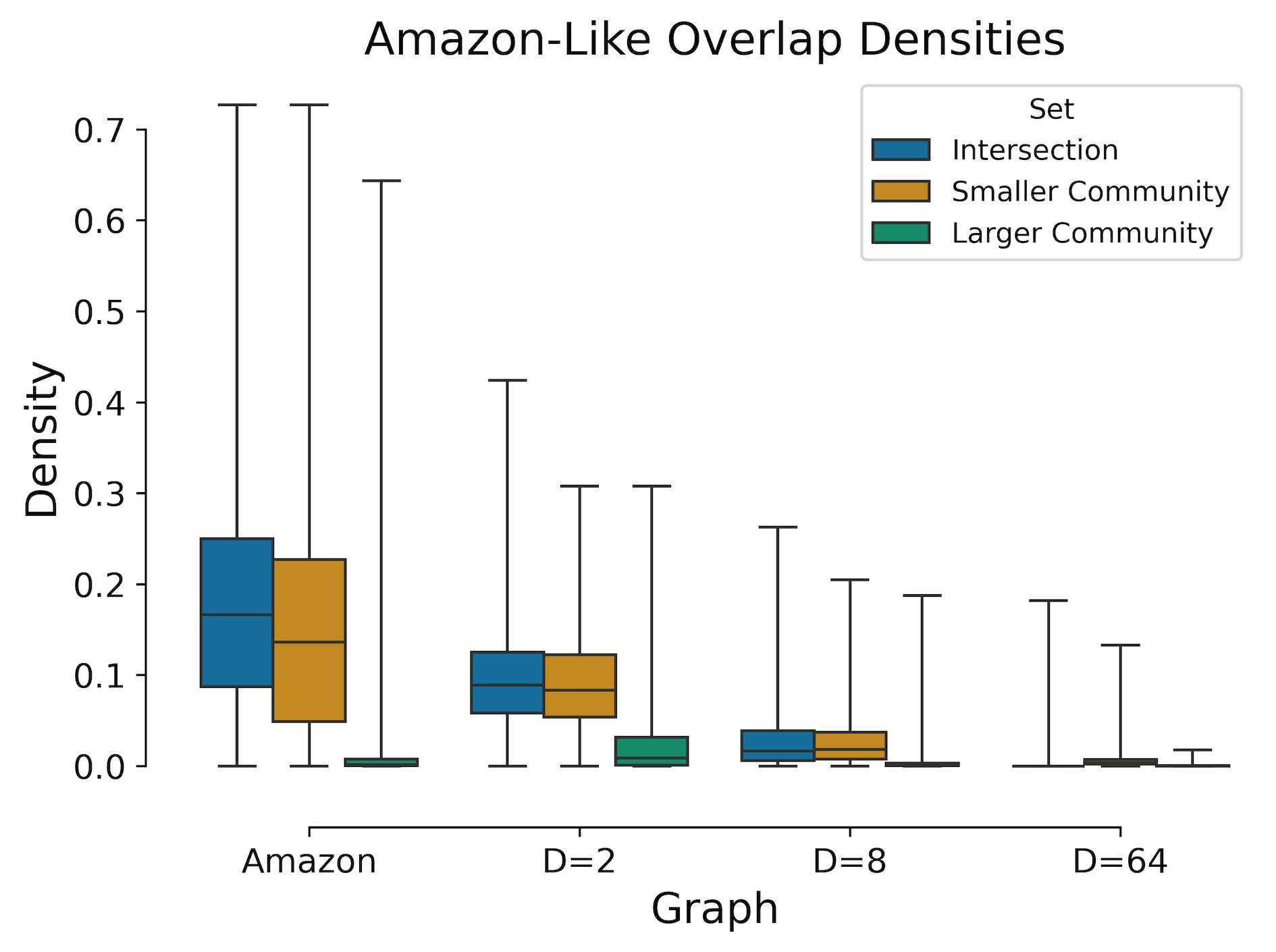}\\
    \includegraphics[width=0.48\linewidth]{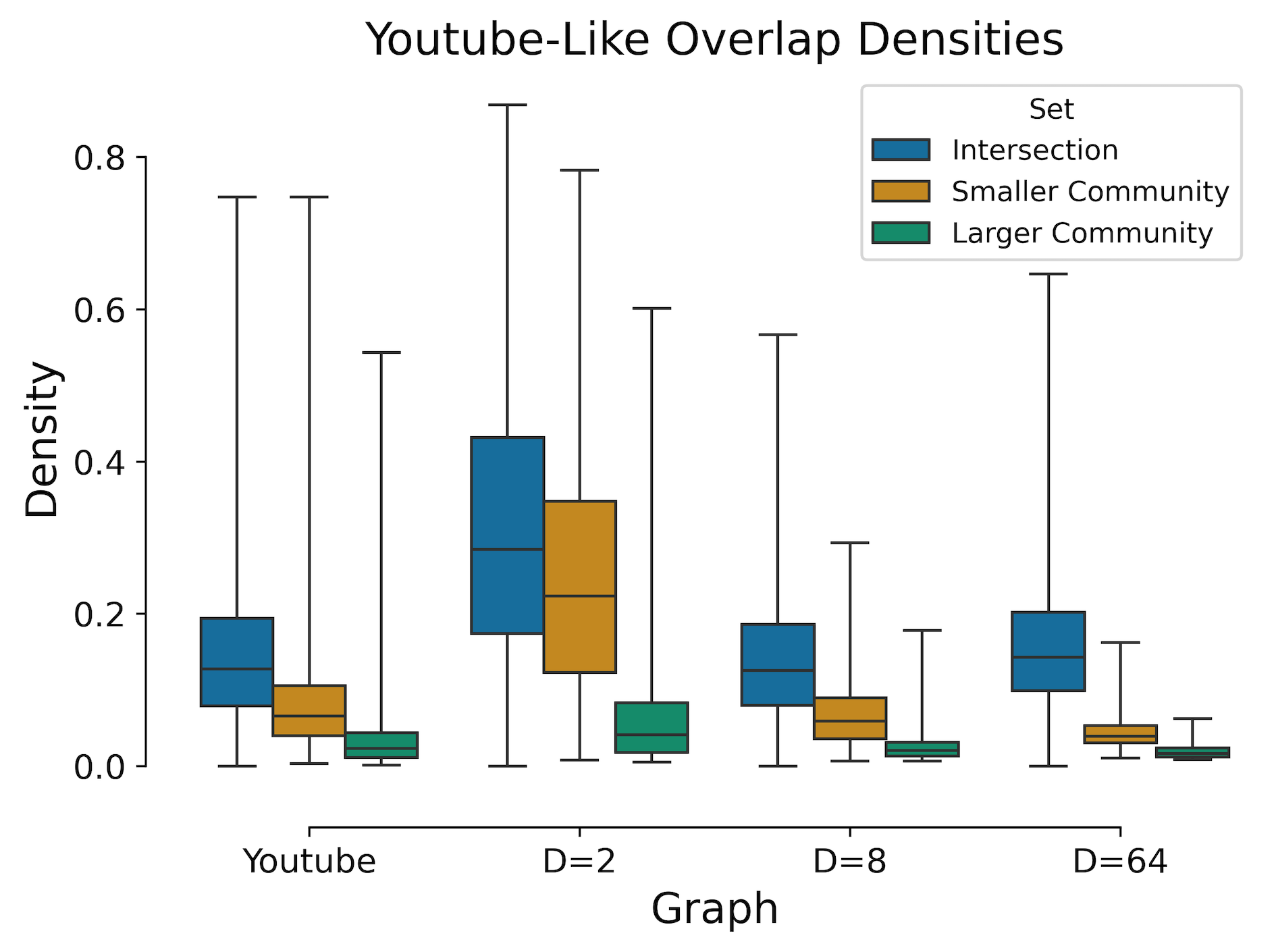}
    \caption{Distribution of the density for pairs of communities with non-zero overlap. The shaded region corresponds to the 25th to 75th quartiles, and the whiskers correspond to the maximum and minimum values. We restrict to overlaps of size at least $10$ due to the discrete and unpredictable nature of small communities.}
    \label{fig:dblp-overlaps}
\end{figure}

Having community intersections that are denser than the respective communities is a desirable property and one of our main motivators for introducing the parameter $\rho$ to the \ABCDoo{} model. In~\cite{yang2014overlapping} it was found that, in real networks, community intersections can be significantly denser than the individual communities. In Figure~\ref{fig:dblp-overlaps} we confirm that the DBLP, Amazon, and YouTube networks all exhibit this phenomenon. Moreover, we compare these densities with those of the corresponding \ABCDoo{} graphs. Aside from confirming that the model can produce behaviour similar to the real networks, we also find that the dimension of the reference layer affects the density of both the communities and the intersections. Moreover, we see that a low-dimensional reference layer yields a better approximation of the real data in the case of Amazon, whereas a high-dimensional reference layer yields a better approximation in the cases of DBLP and YouTube. In the case of DBLP, the difference in distribution between dimensions is partially attributable to the difference in the empirical $\rho$ (recall with Table~\ref{tab:rho} that the low dimension graphs could not achieve a large correlation. Note that the most similar dimension for each real graph is consistent with the experiment presented in Figure~\ref{fig:coms-per-node}.

\clearpage

\subsubsection*{Internal edge fraction}

\begin{figure}[ht]
    \centering
    \includegraphics[width=0.48\linewidth]{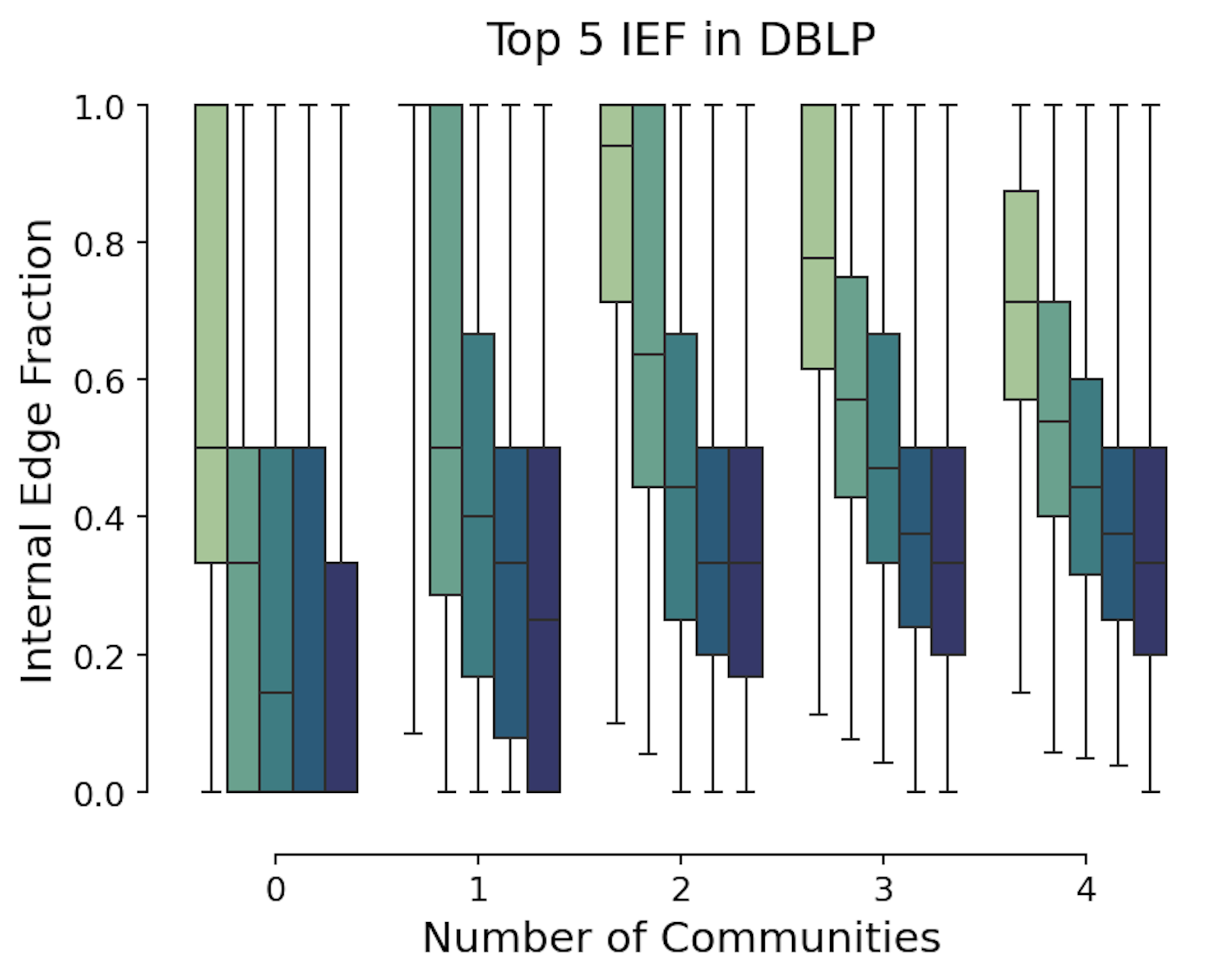}
    \hfill
    \includegraphics[width=0.48\linewidth]{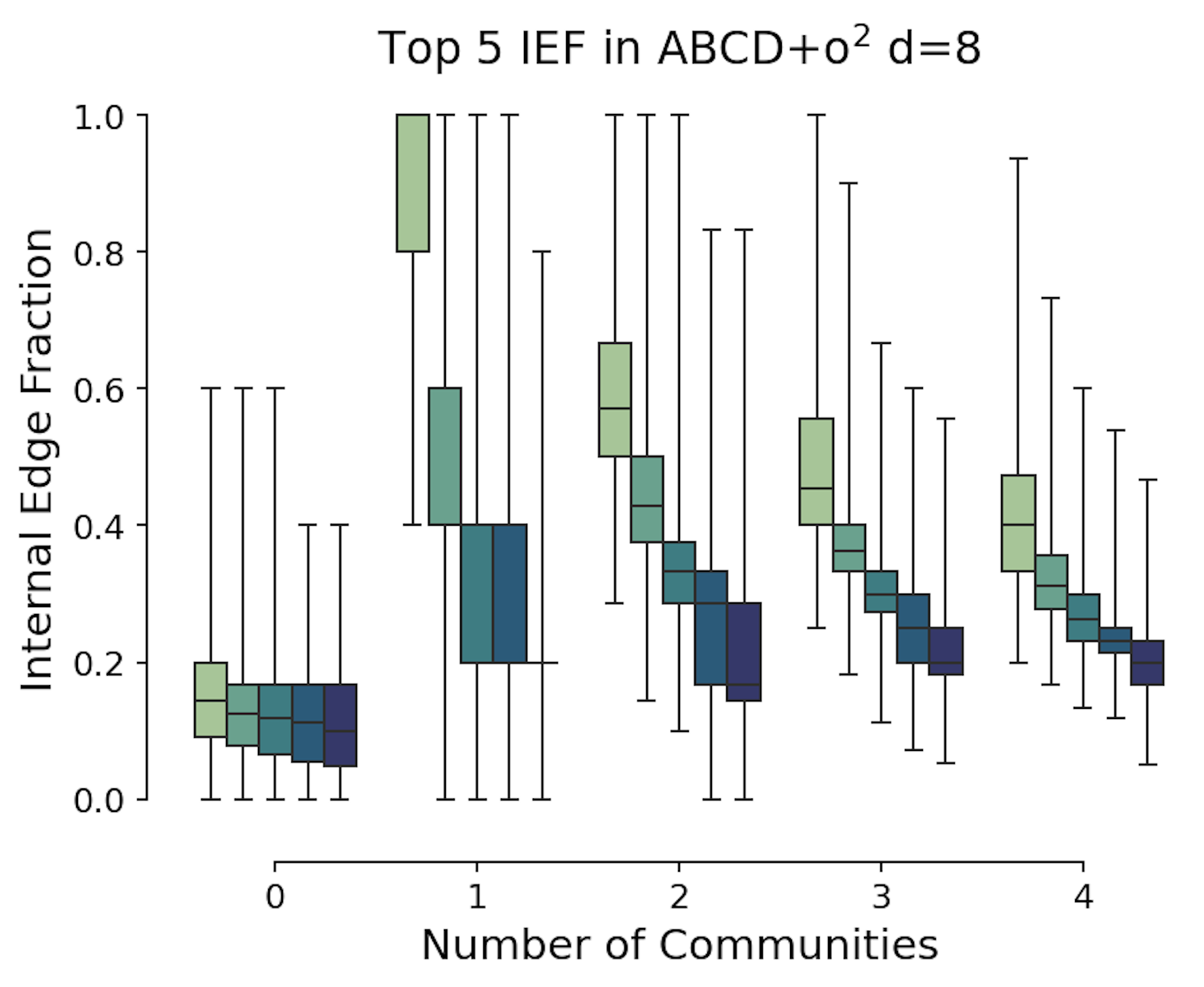}
    \caption{Five largest internal edge fractions, sorted in descending order, grouped by the number of communities. A pattern is consistent between DBLP and \ABCDoo\ , namely, that for nodes belonging to $\ell$ communities there are $\ell$ large, but not equal, internal edge fraction values.}
    \label{fig:dblp-ief}
\end{figure}

Finally, we analyze how strongly nodes are connected to their communities. There are many measures for community association strength, some of which we recently analyzed in~\cite{Betastar2025}. For the sake of simplicity, the measure we will use here is the internal edge fraction $\mathrm{IEF}(v, C)$, defined as

\[
\mathrm{IEF}(v, C) = \frac{|\{ \{u,v\} \in E: u \in C\}|}{\deg(v)} \,.
\]
For a graph $G$, a collection of communities $\{C_i, i \in [k]\}$, and a node $v$, let $C^{(v)}_1, \dots, C^{(v)}_k$ be an ordering of the communities such that 

\[
\mathrm{IEF}(v, C^{(v)}_1) \geq \mathrm{IEF}(v, C^{(v)}_2) \dots \geq \mathrm{IEF}(v, C^{(v)}_k) \,.
\]
Then, if $v$ is a member of $\ell$ communities, we would expect a jump in value between $\mathrm{IEF}(v, C^{(v)}_\ell)$ and $\mathrm{IEF}(v, C^{(v)}_{\ell + 1})$. In Figure~\ref{fig:dblp-ief} we show the range of the top five internal edge fractions, sorted in descending order, on DBLP (left) and a generated \ABCDoo\ graph (right), binned by the number of communities the node $v$ is in. As predicted, a node in $\ell$ communities tends to have $\ell$ relatively large internal edge fractions. Furthermore, despite the \ABCDoo\ model splitting the community degree of a node $v$ evenly among its $\ell$ communities, it is possible for $v$ to have a higher internal edge fraction into community $C$ if either some of $v$'s edges lies in an intersection containing $C$ or some of $v$'s background edges land in $C$. Thus, we see a pattern in \ABCDoo\ graphs where there are $\ell$ large, but not equal, internal edge fraction values. This feature is also present in the DBLP graph, the other real graphs, and the other \ABCDoo\ graphs (see Appendix~\ref{a:ief} for a complete collection of figures).

\section{Benchmarking Community Detection Algorithms}\label{sec:benchmarking}

The main purpose of having synthetic models with ground-truth community structure is to test, tune, and benchmark community detection algorithms. To showcase \ABCDoo\ in this light, we use the model to evaluate the performance of four community detection algorithms. The algorithms are as follows. 

\smallskip
\noindent
\textbf{Leiden~\cite{leiden19}:} a greedy algorithm that attempts to optimize the modularity function. Note that this algorithm returns a partition, and we use it merely as a baseline to compare with algorithms that attempt to find overlapping communities. 

\smallskip
\noindent
\textbf{Edge Clustering~\cite{kim2015detecting}:} an edge-partitioning algorithm that translates to overlapping clusters of nodes. Pairs of edges are measured based on similarity of neighbourhoods, and these similarity measures dictate the order in which edge-communities merge. As edge-communities merge, the modularity is tracked on the line-graph, and the maximum modularity attained yields the edge-communities, which in turn yields overlapping node-communities. 

\smallskip
\noindent
\textbf{Ego-Split~\cite{epasto2017ego}:} a method which finds overlapping clusters in a graph $G$ by applying a partitioning algorithm such as Leiden to an auxiliary graph $G'$ and mapping the resulting partition onto $G$. The auxiliary graph $G'$ is constructed from $G$ by creating multiple copies, or ``egos'', of each node based on its neighbourhood. 

\smallskip
\noindent
\textbf{Ego-Split+IEF~\cite{Betastar2025}:} the same algorithm as Ego-Split, but with a post-processing step that re-assigns nodes to communities based on the IEF measure.

This is by no means an exhaustive list of community detection algorithms. Moreover, we use our own straightforward Python implementations of the above algorithms which may not be optimal. We wish only to showcase the usefulness of \ABCDoo\ in comparing various algorithms related to the community structure of graphs.

\bigskip

In the first experiment, we compare the accuracy of each algorithm with respect to the ground-truth communities. The measure we use to determine accuracy is the overlapping Normalized Mutual Information (oNMI) measure~\cite{mcdaid2011normalized}. For this experiment we fix the number of nodes $n = 5,000$, the minimum degree $\delta = 5$, the maximum degree $\Delta = 100$, the minimum community size $s = 50$, and the maximum community size $S = 500$. The parameters of \ABCDoo{} with the most influence on the quality of detection algorithms are $\xi$ (the level of noise) and $\eta$ (the average number of communities a non-outlier is part of), and we consider every combination of $\xi \in \{0.1, 0.2, \dots , 0.6\}$ and $\eta \in \{1, 1.25, \dots, 2.5\}$. We fix the remaining parameters according to the YouTube graph, namely, $\gamma = 1.87$, $\beta = 2.13$, and $\rho = 0.37$. 

\begin{figure}[ht]
    \centering
    \includegraphics[width=14cm]{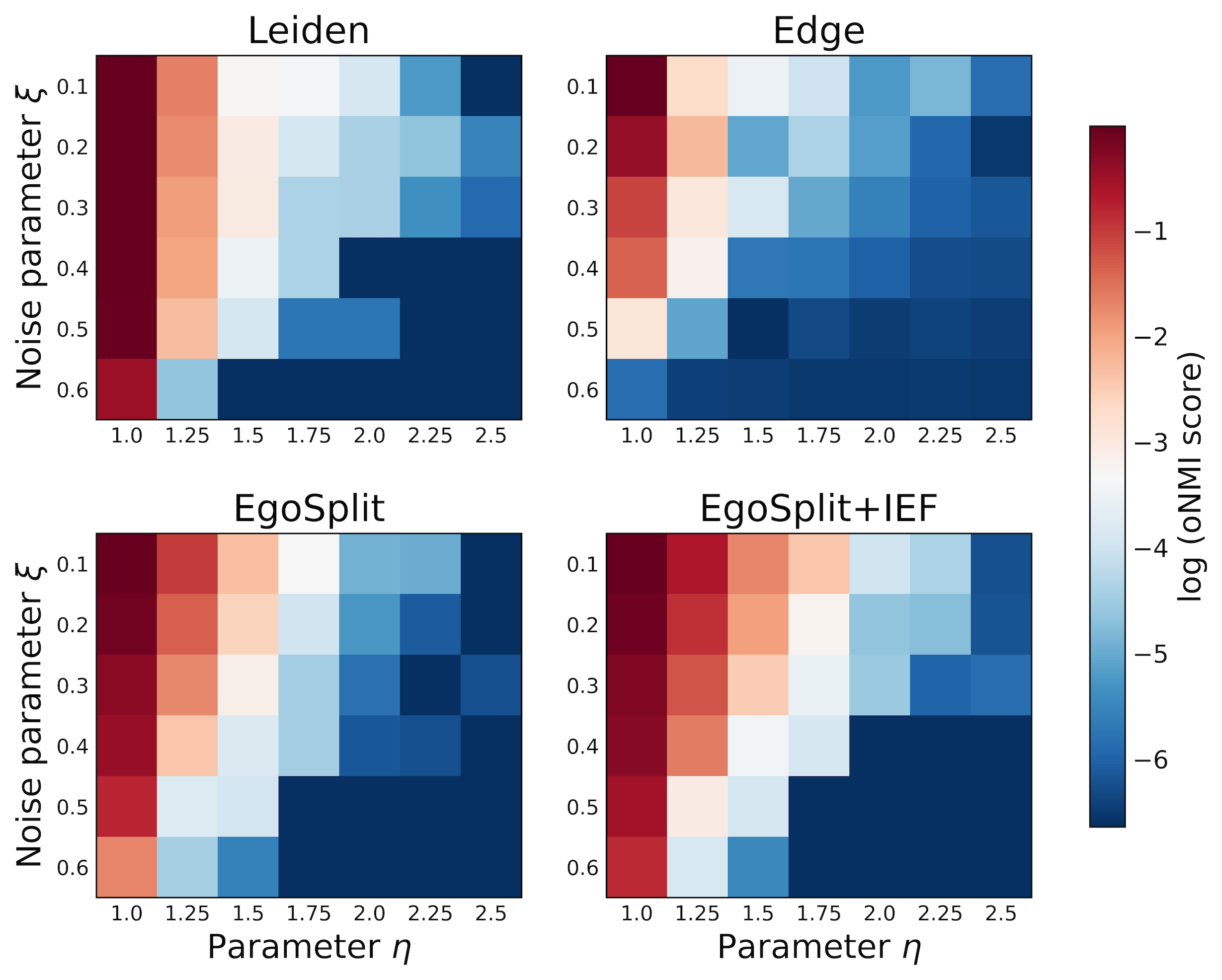}
    \caption{Comparing oNMI of four clustering algorithms on \ABCDoo{} graphs with 5,000 nodes.}
    \label{fig:compare_algos_grid}
\end{figure}

\begin{figure}[ht]
    \centering
    \includegraphics[width=8cm]{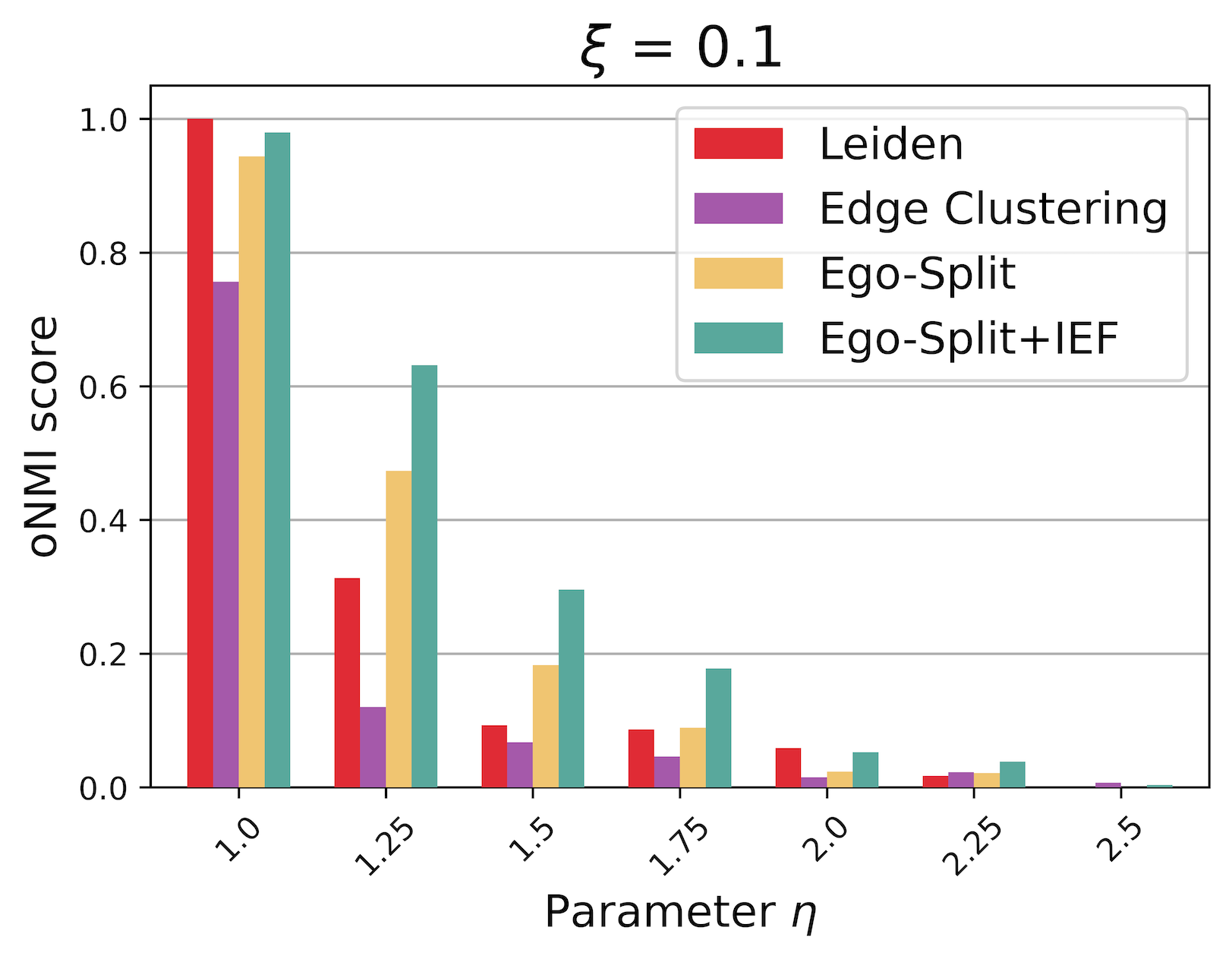}
    \includegraphics[width=8cm]{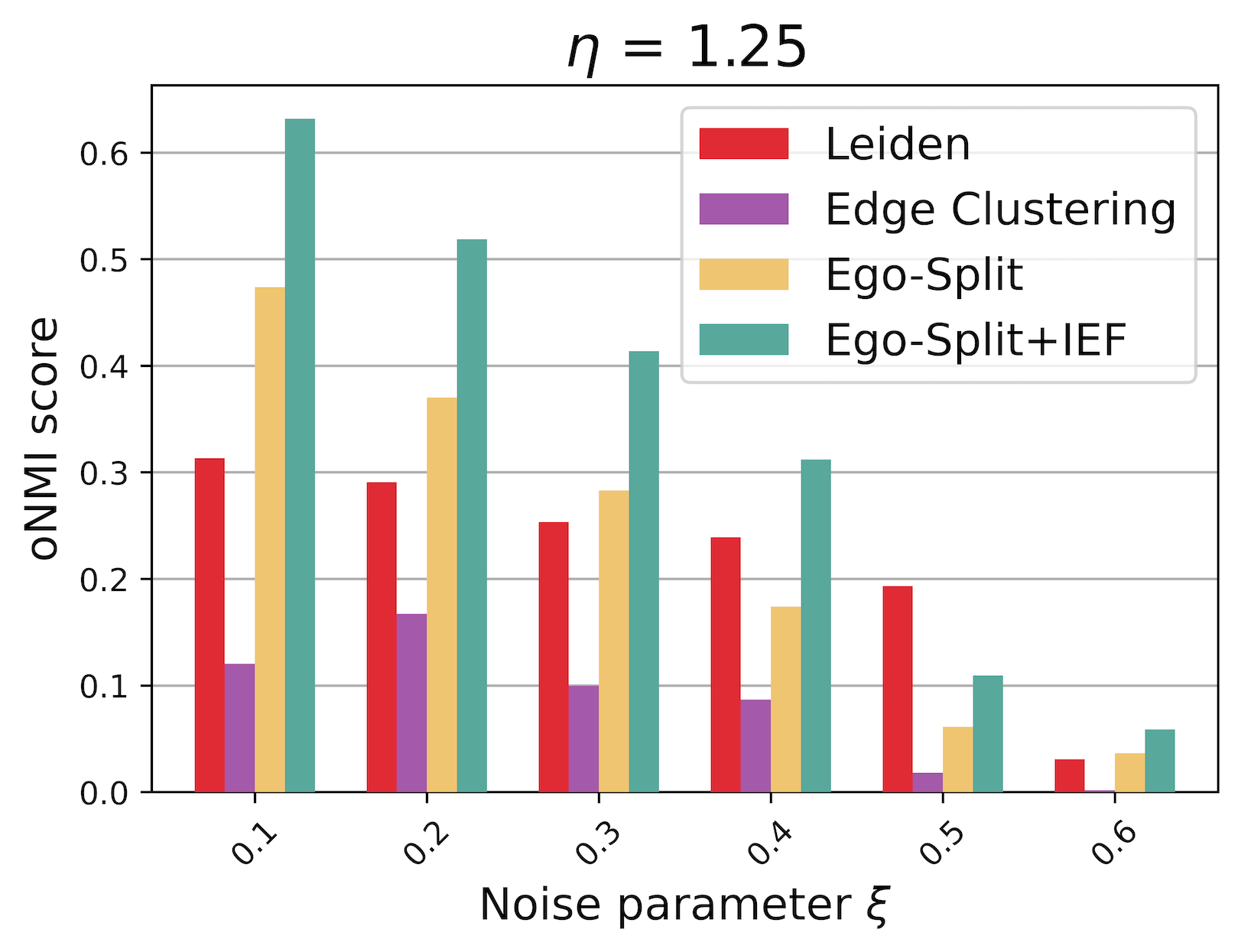}
    \caption{Comparing oNMI of four clustering algorithms on \ABCDoo{} graphs with 5,000 nodes. In the left plot, we fix $\eta=1.25$ and vary $\xi$. In the right plot, we fix $\xi = 0.1$ and vary $\eta$.}
    \label{fig:compare_algos}
\end{figure}

Figure~\ref{fig:compare_algos_grid} summarizes the results of the experiment. We see that the \textbf{Leiden} algorithm performs the best when $\eta = 1$, whereas for $\eta>1$ the \textbf{Ego-Split+IEF} algorithm performs better. Given that the \textbf{Leiden} algorithm finds a partition of the nodes, it is unsurprising that it quickly depreciates in quality as $\eta$ increases. We also see a general trend of all algorithms performing worse as the graph gets noisier, either by increasing $\xi$ or $\eta$. This trend is highlighted separately in Figure~\ref{fig:compare_algos} where we consider only results with $\eta=1.25$ (left) or $\xi=0.1$ (right). From numerous and varying tests, we have found that, in general, increasing $\eta$ is far more damming to community detection algorithms than increasing $\xi$.

\begin{figure}[ht]
    \centering
    \includegraphics[width=14cm]{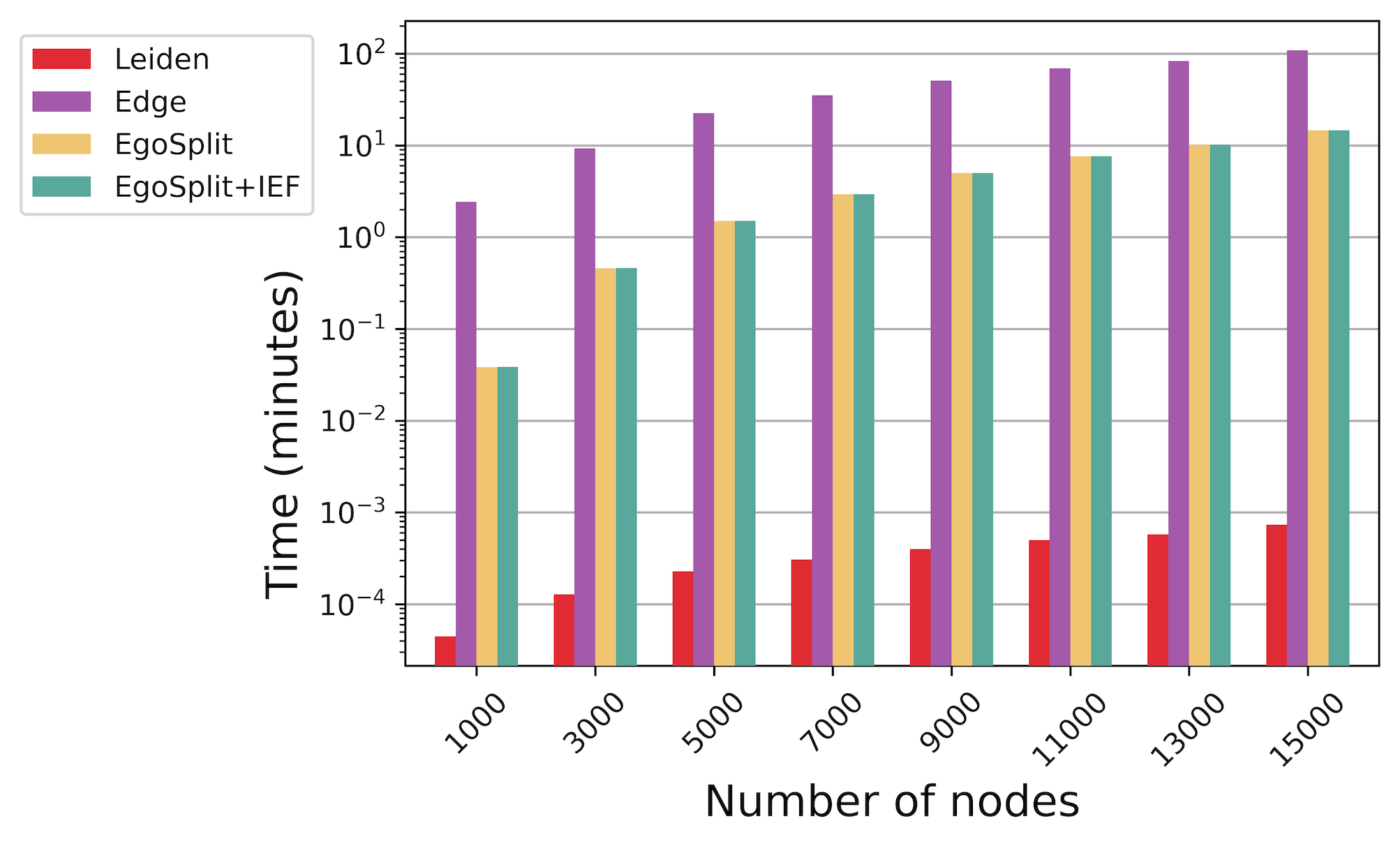}
    \caption{The runtime of four clustering algorithms on \ABCDoo{} graphs with 1,000 to 15,000 nodes.}
    \label{fig:compare_algos_time}
\end{figure}

Another important aspect of clustering algorithms is their performance when running on large graphs. In our second experiment, we fix $\eta=1.25$ and $\xi=0.1$ and we vary the number of nodes from $n=1,000$ to $n=15,000$; the parameters other than $n, \eta$ and $\xi$ are the same as in the previous experiment. The results are presented in Figure~\ref{fig:compare_algos_time}. We see that all algorithms trying to recover overlapping communities are significantly slower than the simple \textbf{Leiden} partitioning algorithm. We also see that adding the IEF-based post-processing stage to the Ego-Splitting algorithm is negligible in terms of performance. Together with the experiment presented in Figure~\ref{fig:compare_algos_grid}, this negligible slowdown suggests that including a post-processing stage to an overlapping community detection algorithm yields a worthwhile improvement. 

\section{Conclusion}\label{sec:conclusion}

We presented \ABCDoo: a generalization of the \textbf{ABCD+o} model that allows for overlapping communities. We proposed a novel mechanism for producing overlapping communities, a hidden reference layer, that requires two new parameters: the average number of communities per node $\eta$, and the dimension for the reference layer $d$. Furthermore, we introduced a parameter $\rho$ that controls the correlation between degree and number of communities for non-outlier nodes. In a series of experiments, we compared the properties of the \ABCDoo\ model to those of real graphs using parameters measured from the real graph. We found that the hidden reference layer can accurately create a structure of overlapping communities similar to those in the real graph. We also found that the model can successfully create an empirical $\rho$ close to the requested value (except for extreme cases). Finally, we showcased the model's ability to benchmark community detection algorithms and compare their quality. 

This paper acts as a first step in our study of the \ABCDoo\ model and gives way to several open problems for future research. First, a theoretical description of the distribution of the number of communities per node and the effect of the dimension of the reference layer are clear goals. A similar direction is understanding the effects of modifying the geometry of the reference layer. For example, one could estimate the reference space of a network from auxiliary data and use it for a corresponding parameter-fitted \ABCDoo\ model. Using this fitting procedure, can we achieve additional structural similarities to real networks?

We are also interested in results for the \ABCDoo\ model that generalize results of the \textbf{ABCD} and \textbf{ABCD+o} models. In~\cite{kaminski2022modularity} the modularity was studied and it was found that the maximum modularity came from the ground truth communities until a certain level of noise, after which a higher modularity could be attained. One could test for a similar result in \ABCDoo\ by comparing the maximum modularity to both the level of noise $\xi$ and the overlap parameter $\eta$. A first step to this future research would be selecting an appropriate generalization of modularity for overlapping sets, such as the generalization proposed in~\cite{devi2016overlapping}. Additionally, in~\cite{self-similarityABCD} it was shown that \textbf{ABCD} graphs exhibit self-similar behaviour, namely, the degree distributions of communities are asymptotically the same as the degree distribution of the whole graph (up to an appropriate normalization). Another open problem is checking if this self-similar property persists in \ABCDoo. 

\bibliography{ESCBib}

\newpage

\appendix
\section{Density of Intersection Figures} \label{a:intersection_density}

\begin{figure}[ht]
    \centering
    \includegraphics[width=0.48\linewidth]{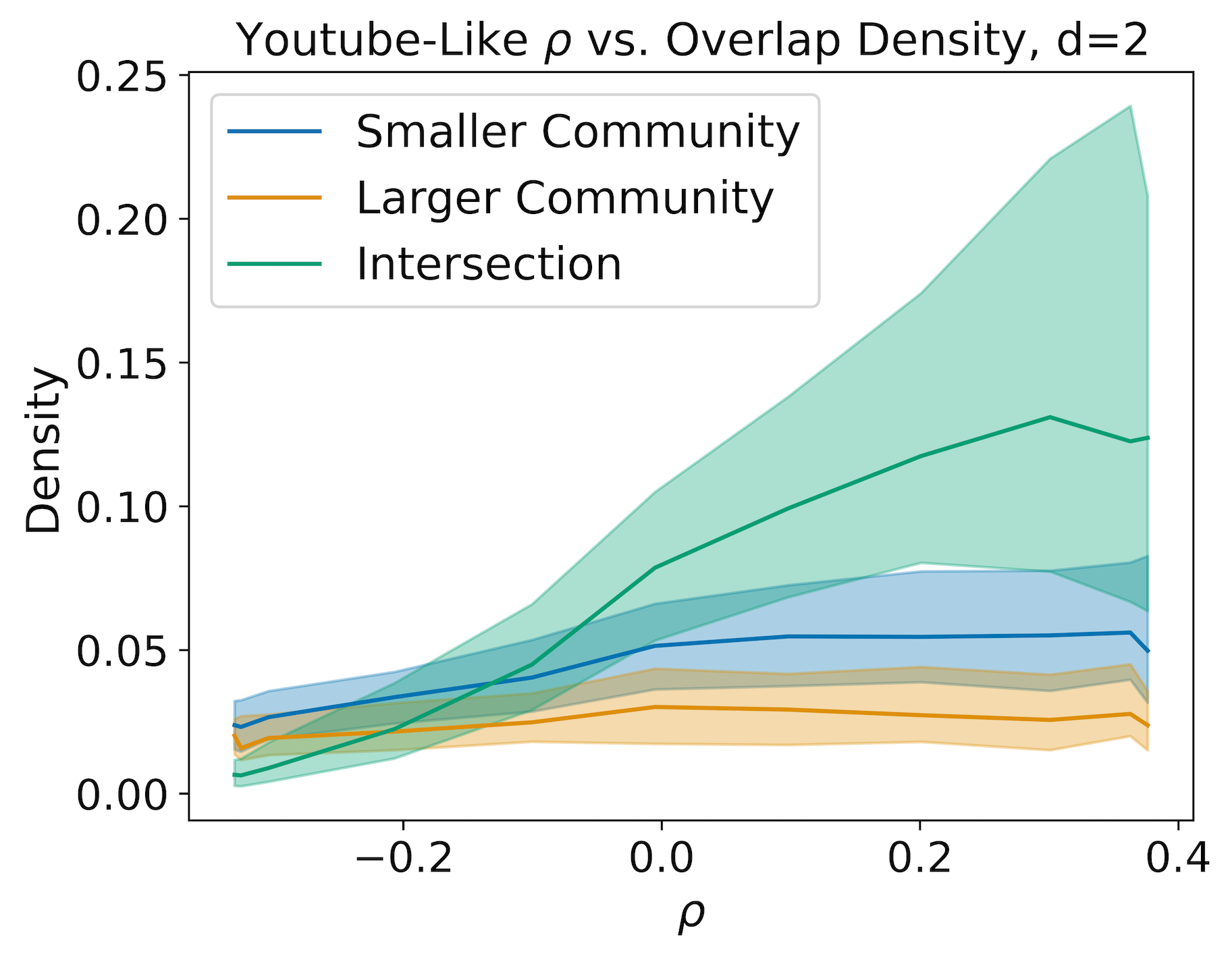} \hfill
    \includegraphics[width=0.48\linewidth]{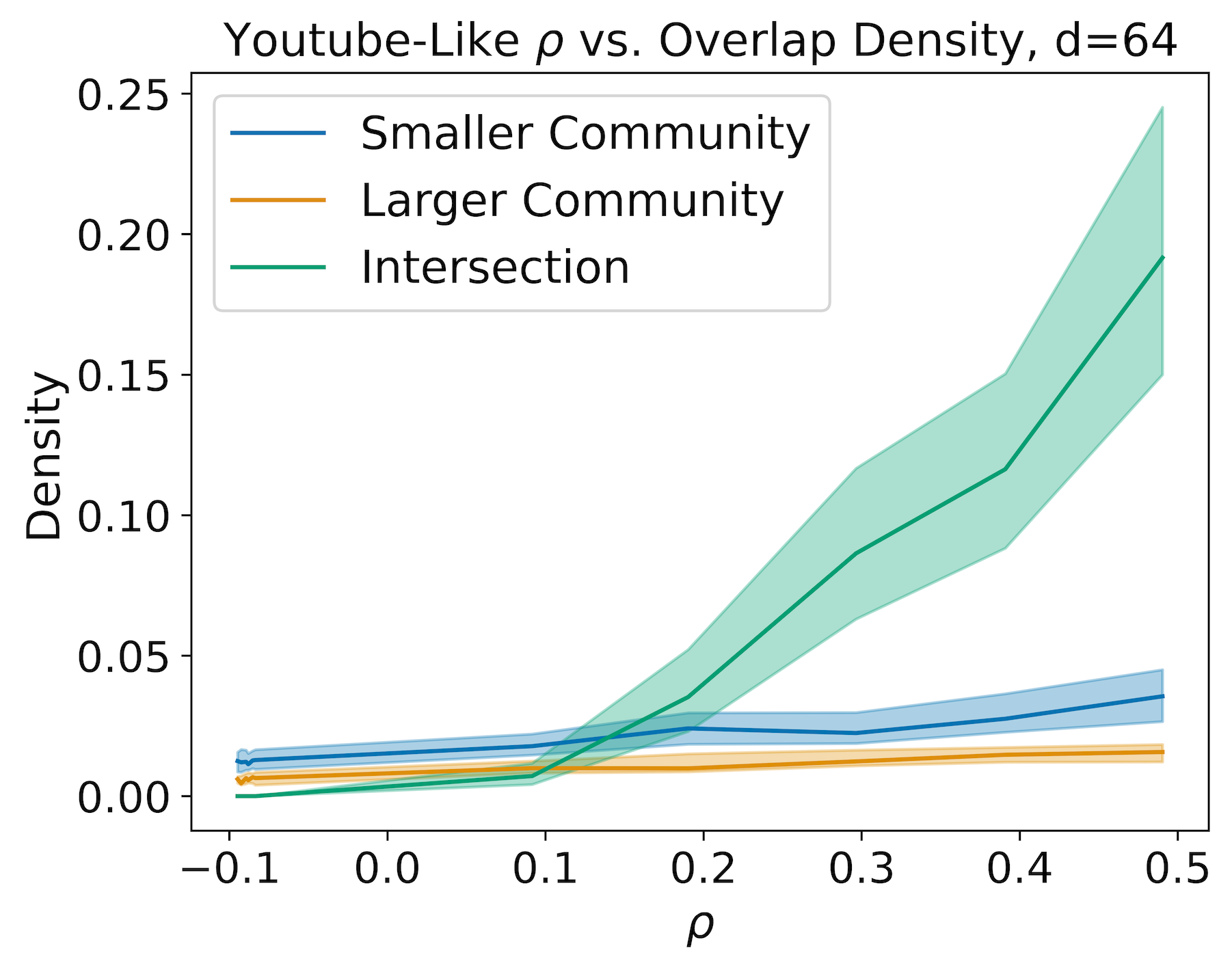}
    \caption{The density of overlaps compared to the empirical $\rho$ on YouTube-like \ABCDoo\ graphs.}
    \label{fig:a1}
\end{figure}

\begin{figure}[ht]
    \centering
    \includegraphics[width=0.48\linewidth]{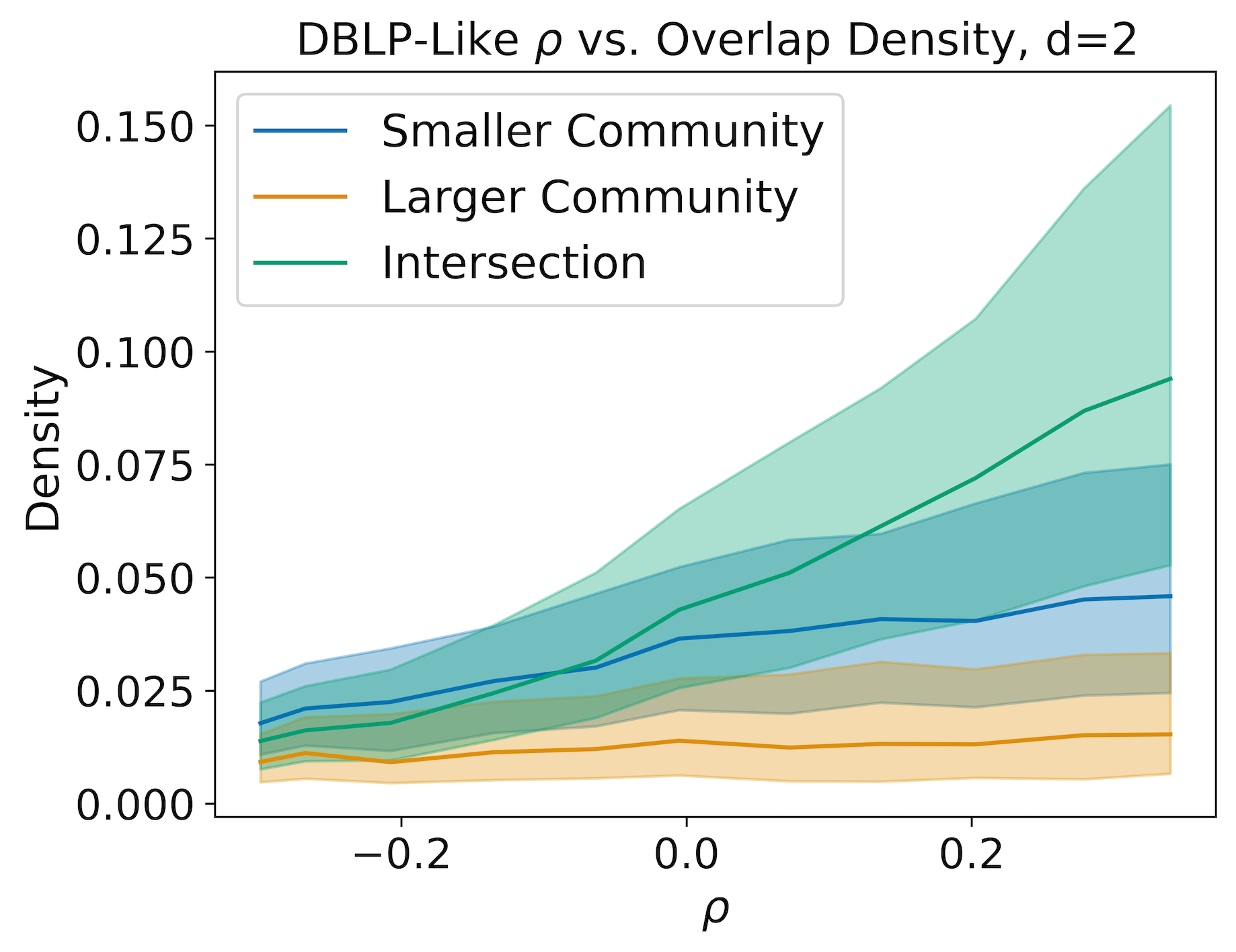} \hfill
    \includegraphics[width=0.48\linewidth]{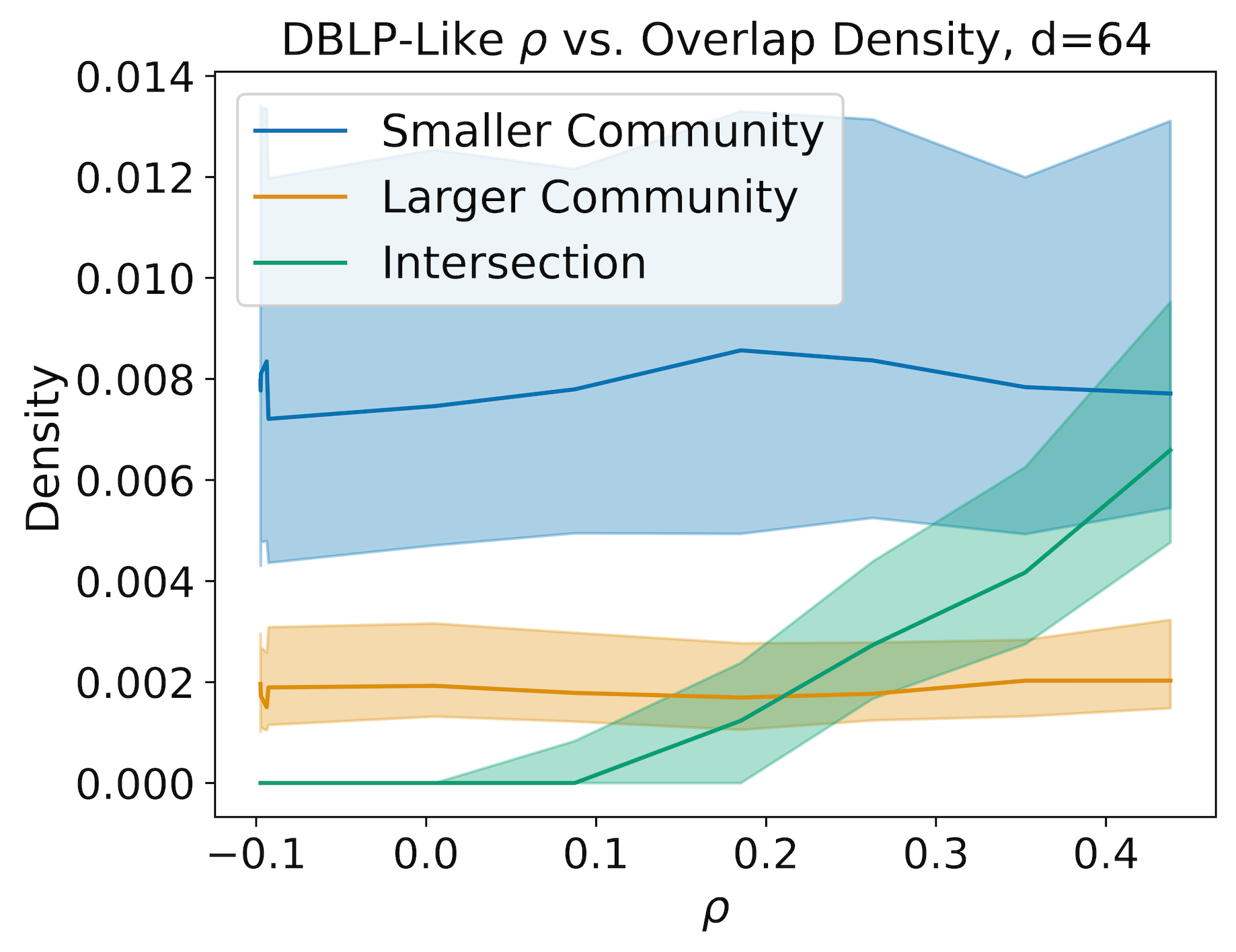}
    \caption{The density of overlaps compared to the empirical $\rho$ on DBLP-like \ABCDoo\ graphs.}
    \label{fig:a2}
\end{figure}

\begin{figure}[ht]
    \centering
    \includegraphics[width=0.48\linewidth]{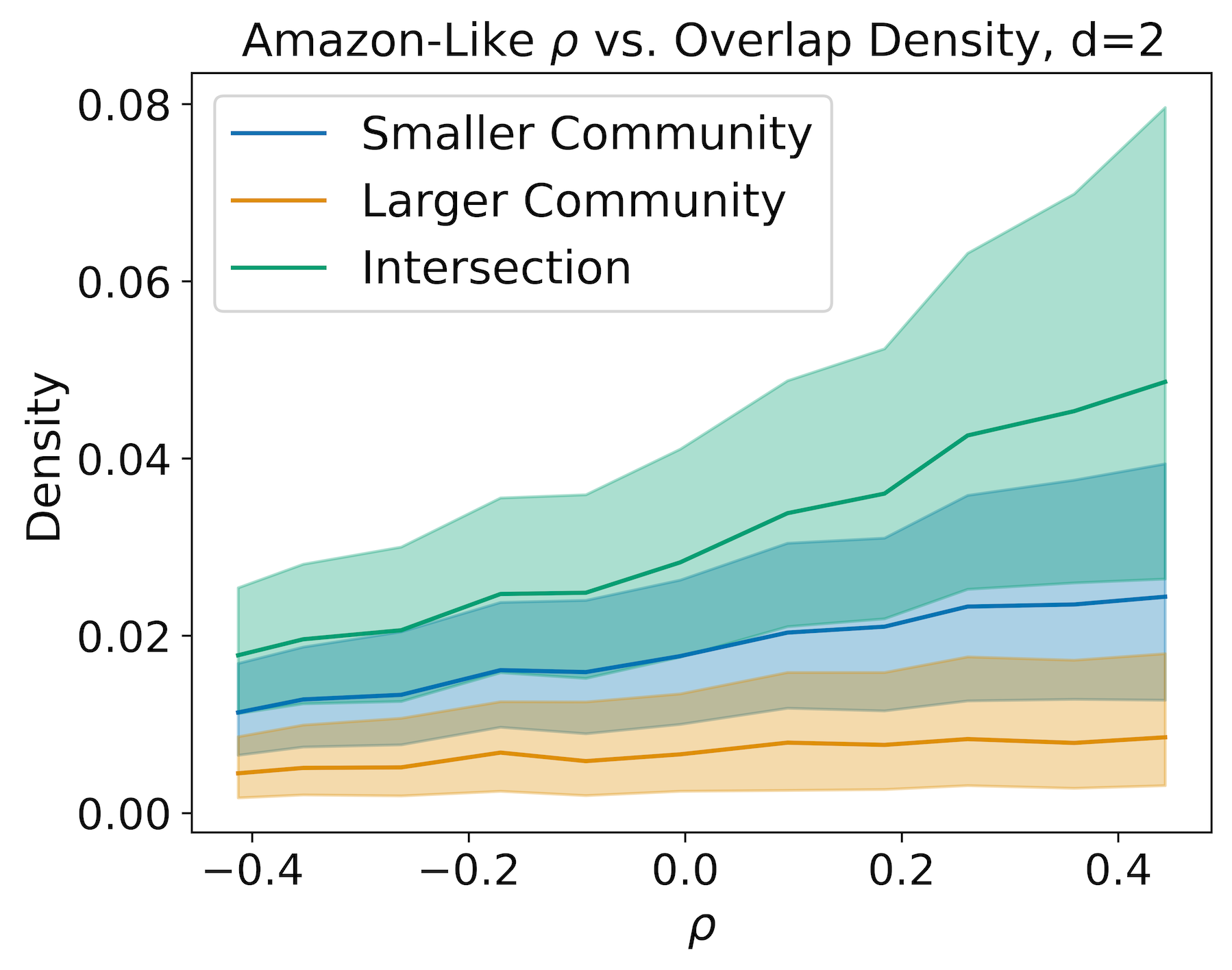}
    \includegraphics[width=0.48\linewidth]{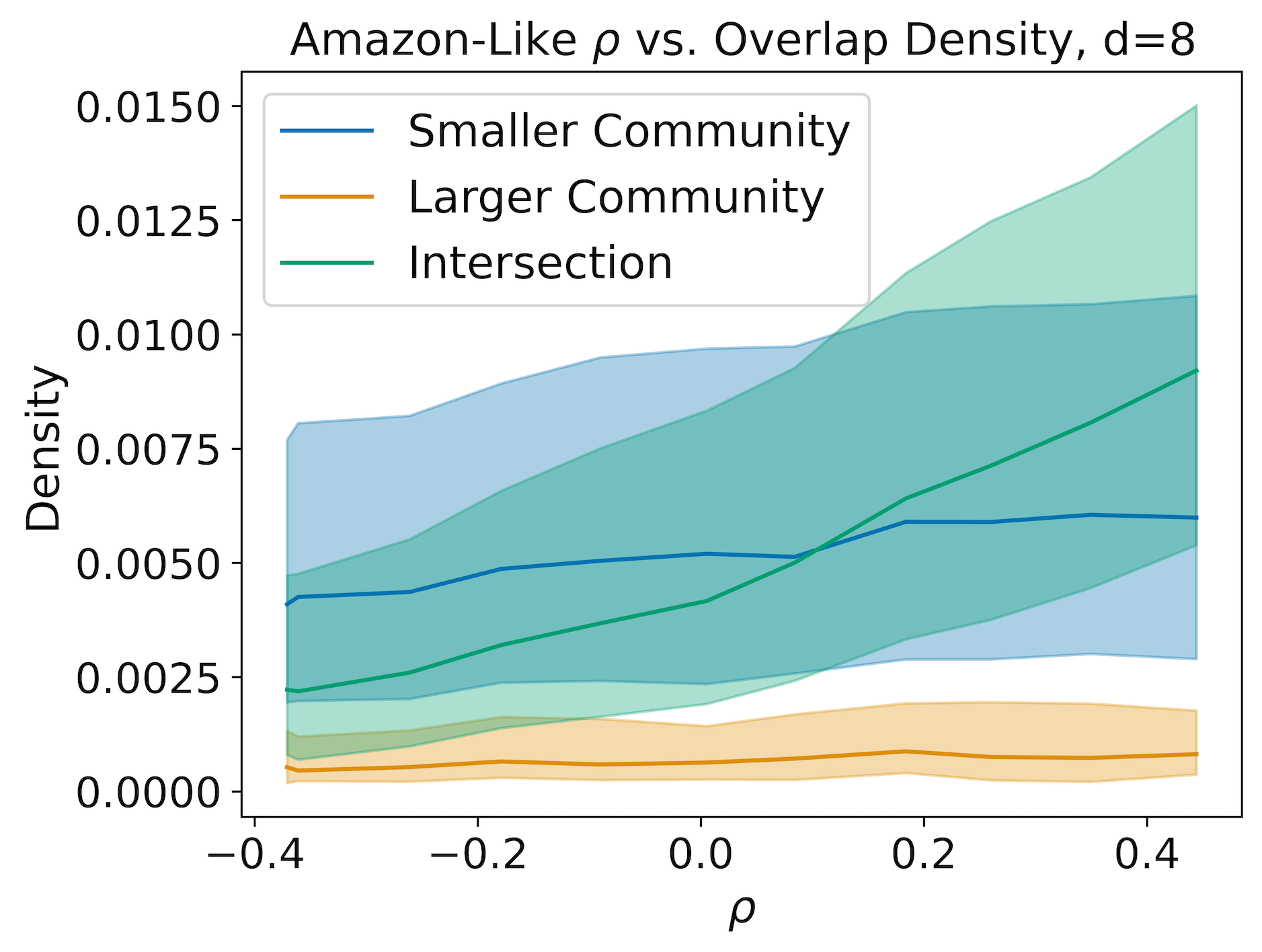}\\
    \includegraphics[width=0.48\linewidth]{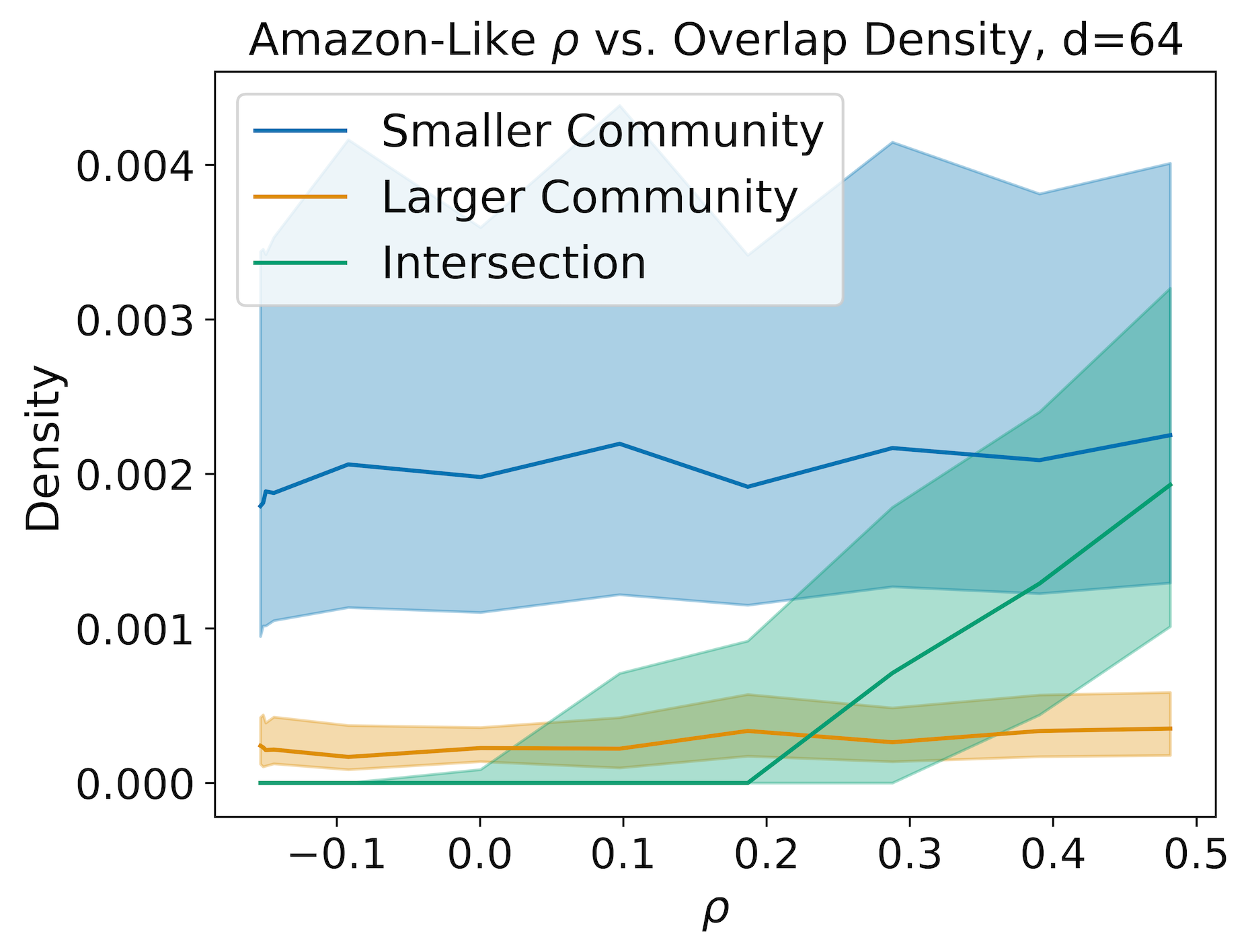}
    \caption{The density of overlaps compared to the empirical $\rho$ on Amazon-like \ABCDoo\ graphs.}
    \label{fig:a3}
\end{figure}

Figures~\ref{fig:a1},~\ref{fig:a2} and~\ref{fig:a3}, together with the previously shown Figure~\ref{fig:intersection density vs rho}, present the full experiment comparing community and intersection densities for various $\rho$.

\bigskip

\clearpage 

\section{Internal Edge Fraction Figures} \label{a:ief}

\begin{figure}[ht]
    \centering
    \includegraphics[width=0.48\linewidth]{dblp_ief.png}\hfill
    \includegraphics[width=0.48\linewidth]{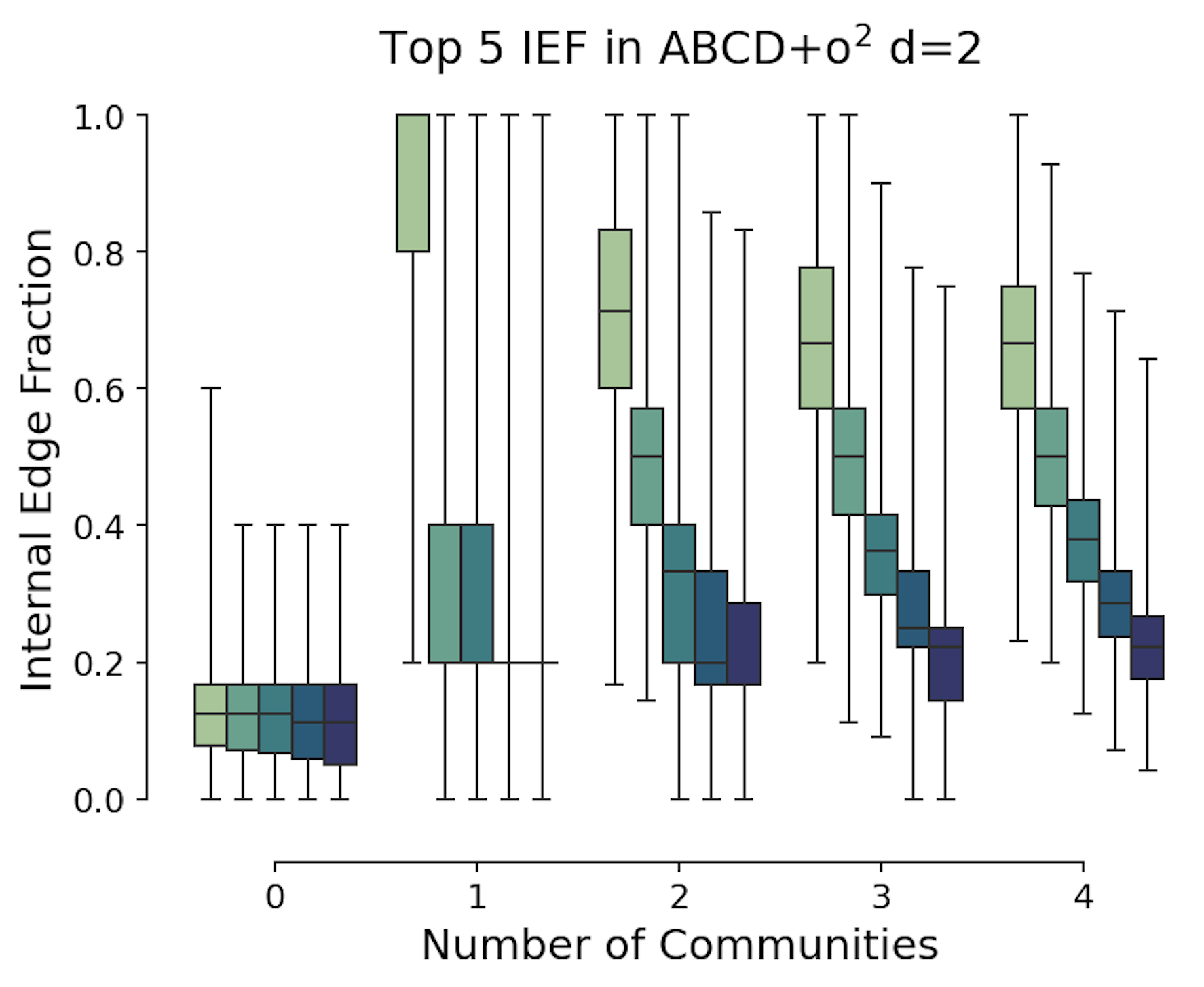}\\\vspace{3em}
    \includegraphics[width=0.48\linewidth]{dblp_ief_d8.png}\hfill
    \includegraphics[width=0.48\linewidth]{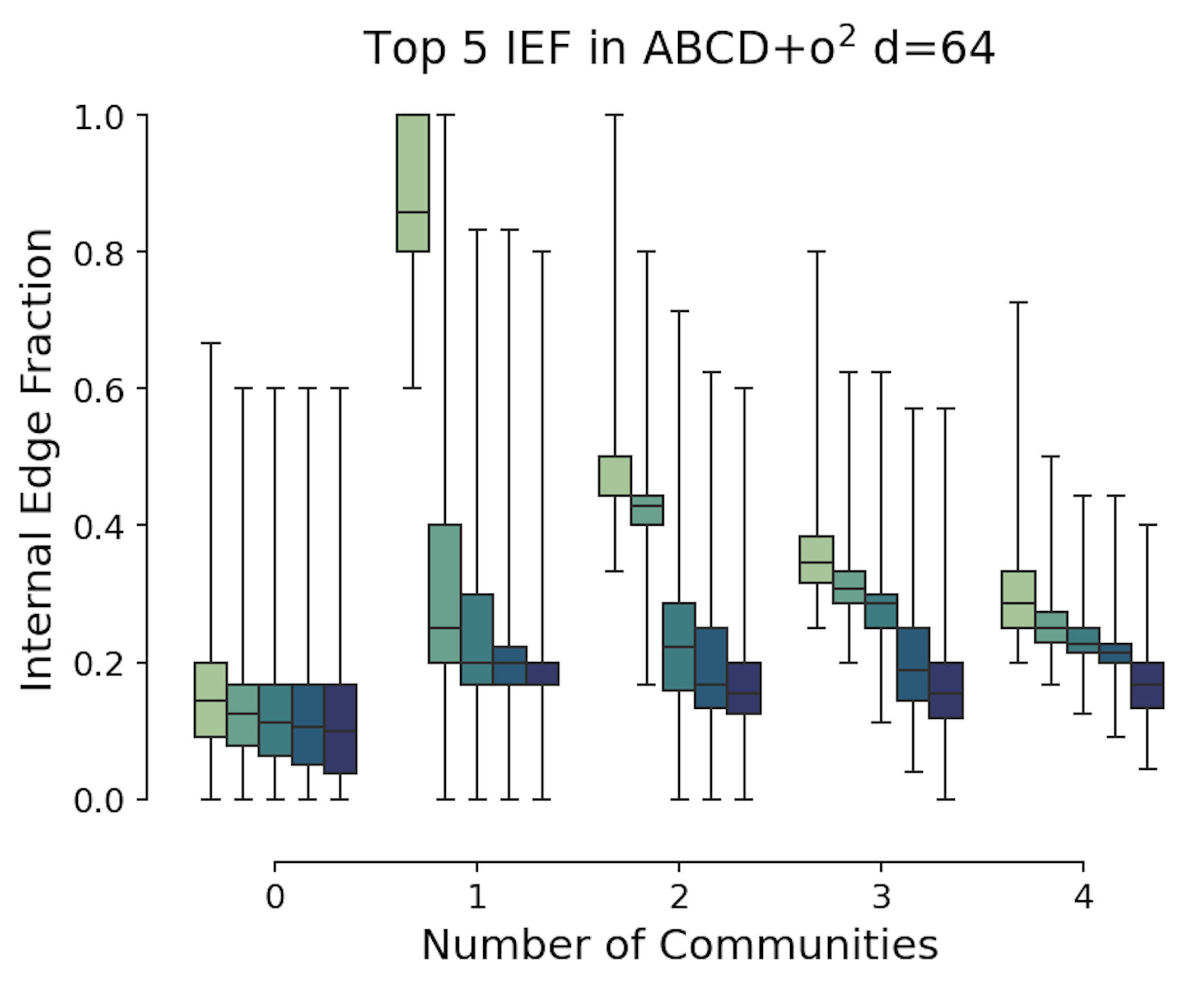}
    \caption{The top 5 IEF values in decreasing order, grouped by the number of communities the node belongs to, for DBLP and DBLP-like \ABCDoo graphs.}
    \label{fig:b1}
\end{figure}

\begin{figure}[ht]
    \centering
    \includegraphics[width=0.48\linewidth]{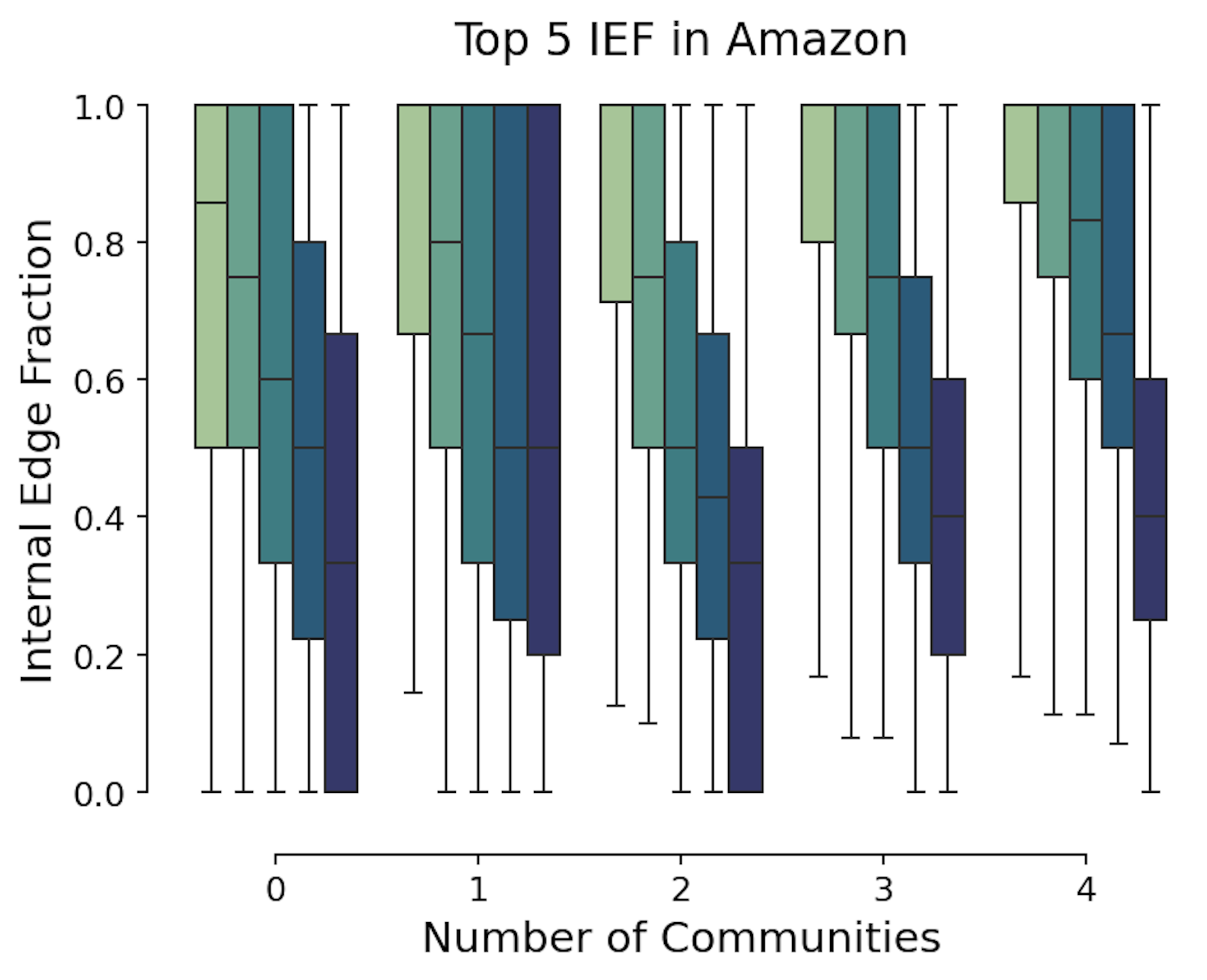}\hfill
    \includegraphics[width=0.48\linewidth]{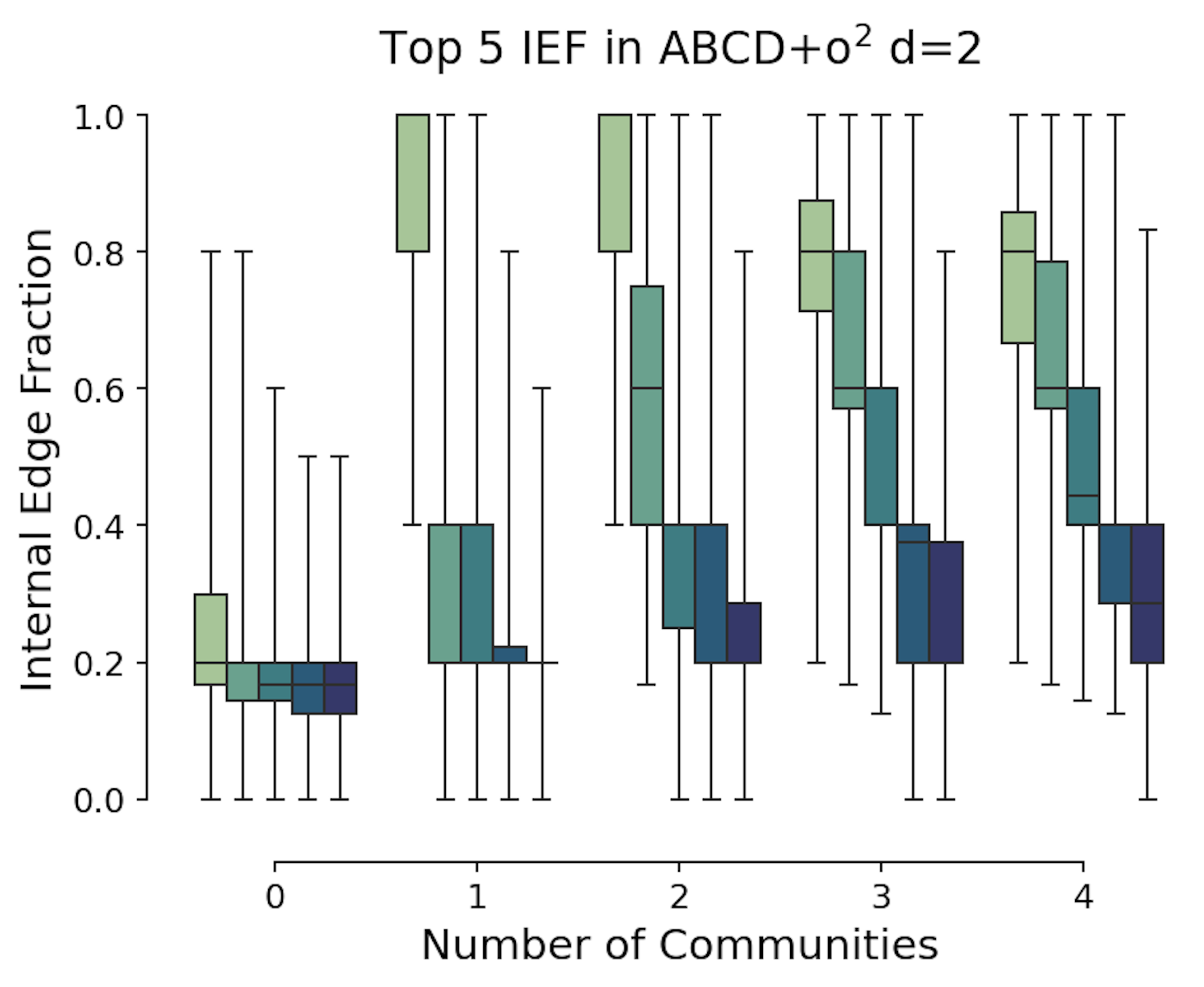}\\\vspace{3em}
    \includegraphics[width=0.48\linewidth]{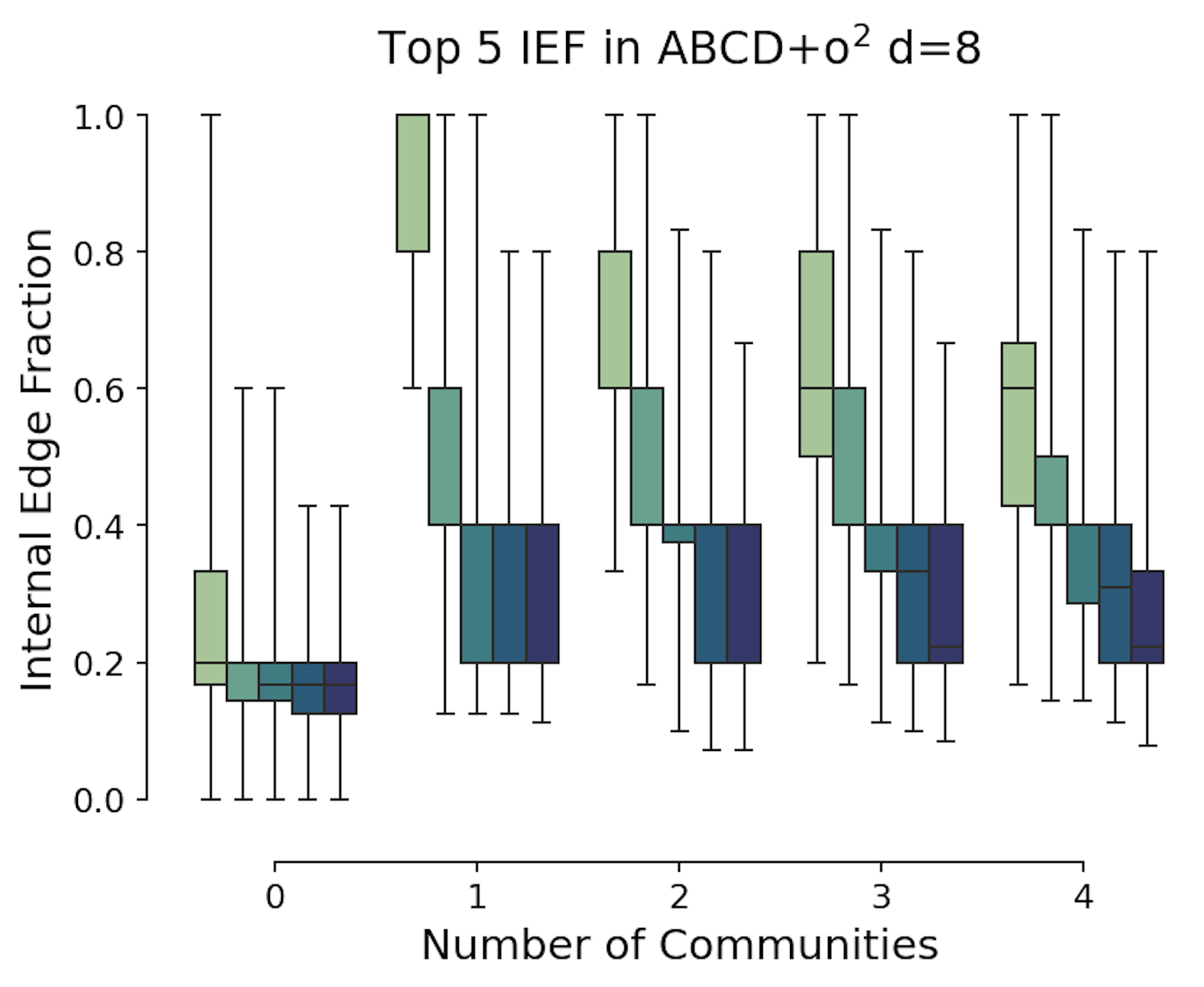}\hfill
    \includegraphics[width=0.48\linewidth]{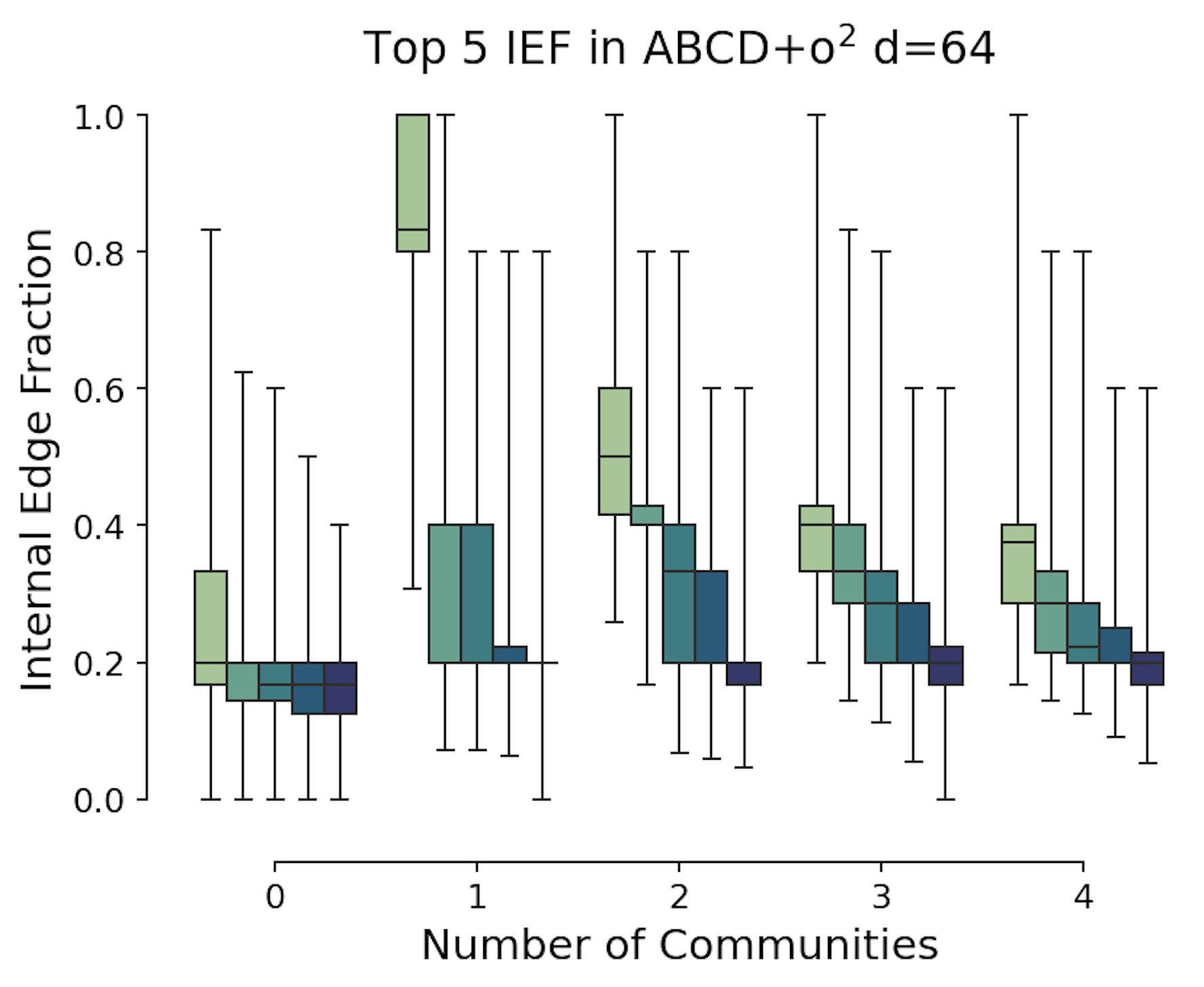}
    \caption{The top 5 IEF values in decreasing order, grouped by the number of communities the node belongs to, for Amazon and Amazon-like \ABCDoo graphs.}
    \label{fig:b2}
\end{figure}

\begin{figure}[ht]
    \centering
    \includegraphics[width=0.48\linewidth]{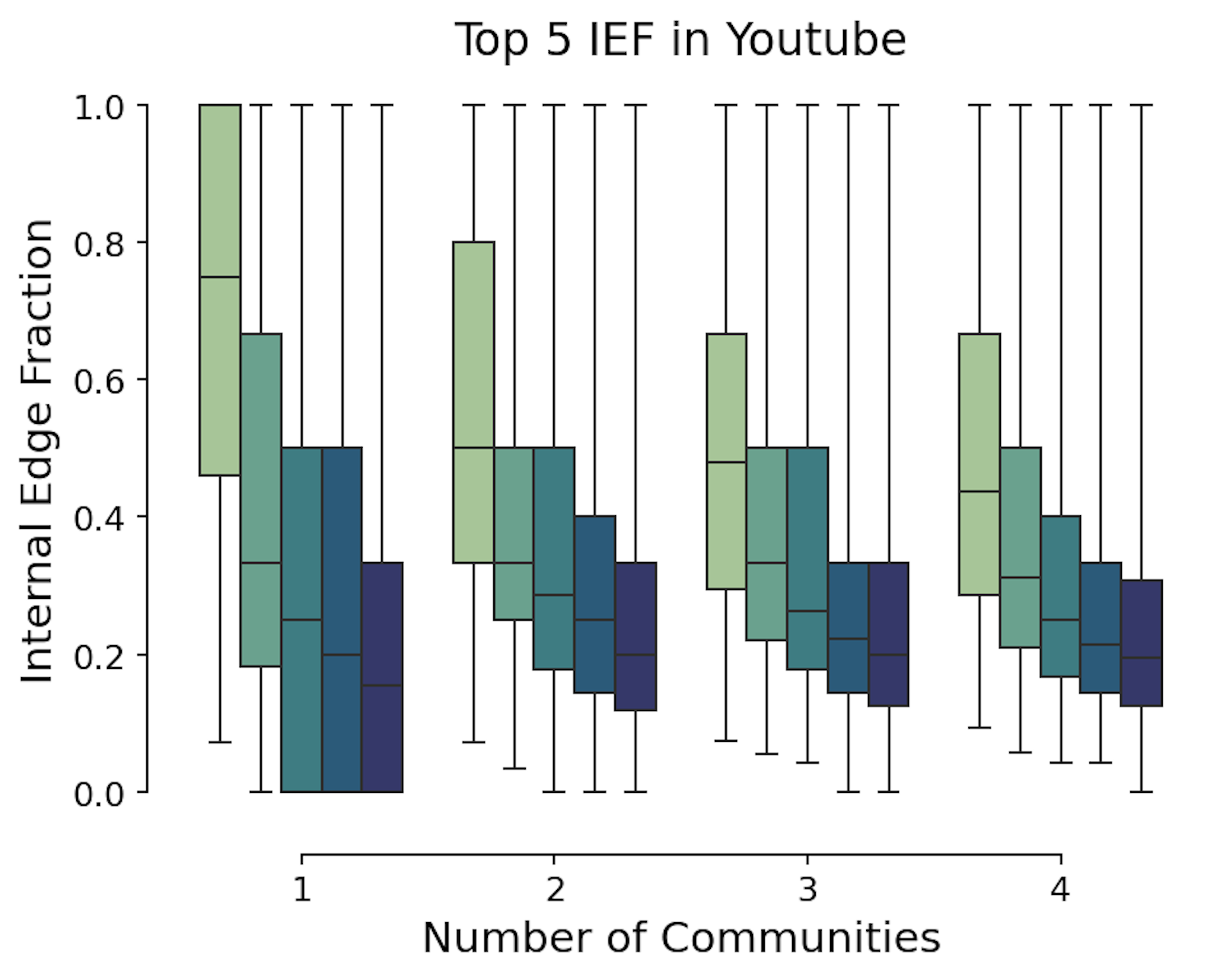}\hfill
    \includegraphics[width=0.48\linewidth]{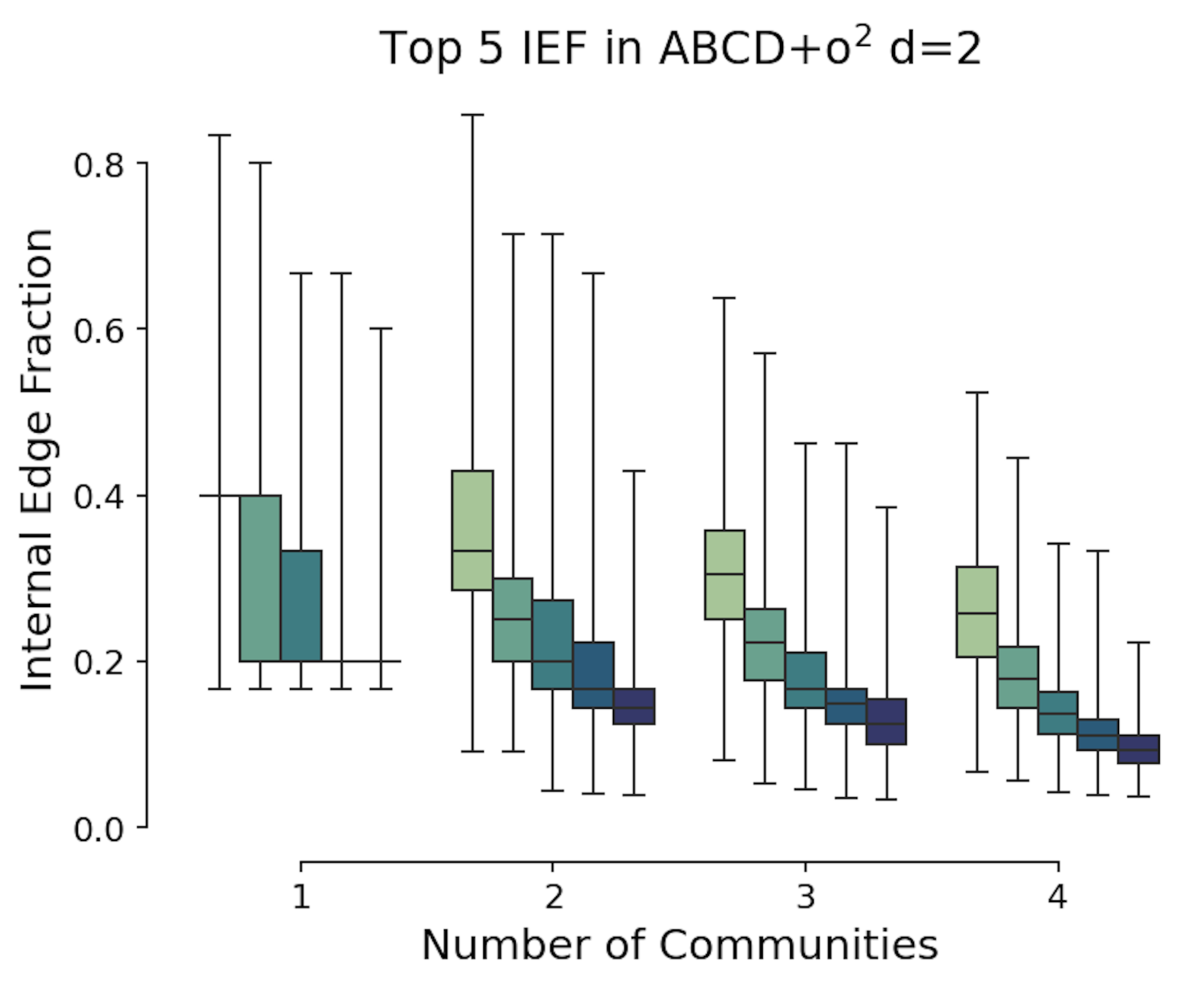}\\\vspace{3em}
    \includegraphics[width=0.48\linewidth]{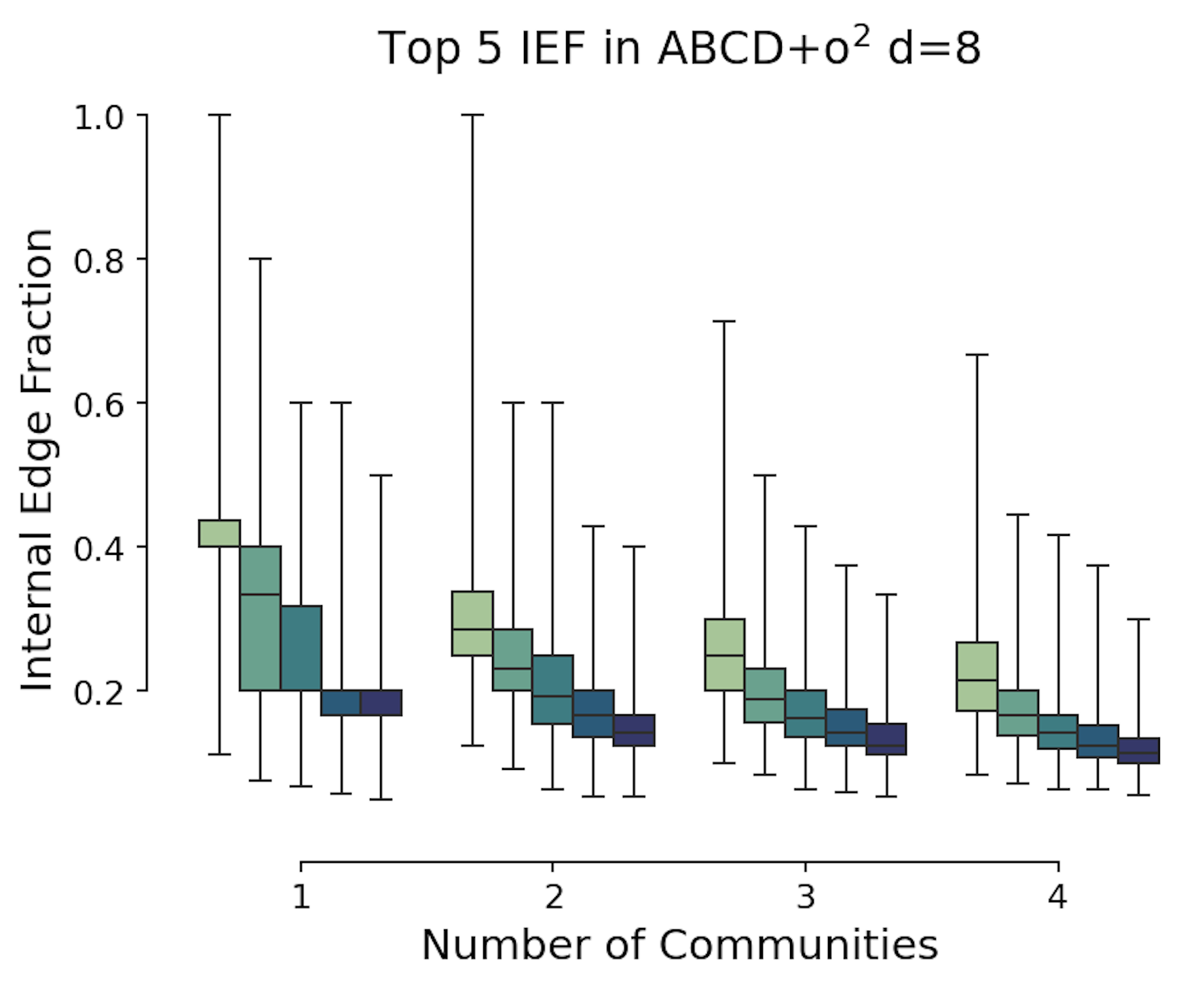}\hfill
    \includegraphics[width=0.48\linewidth]{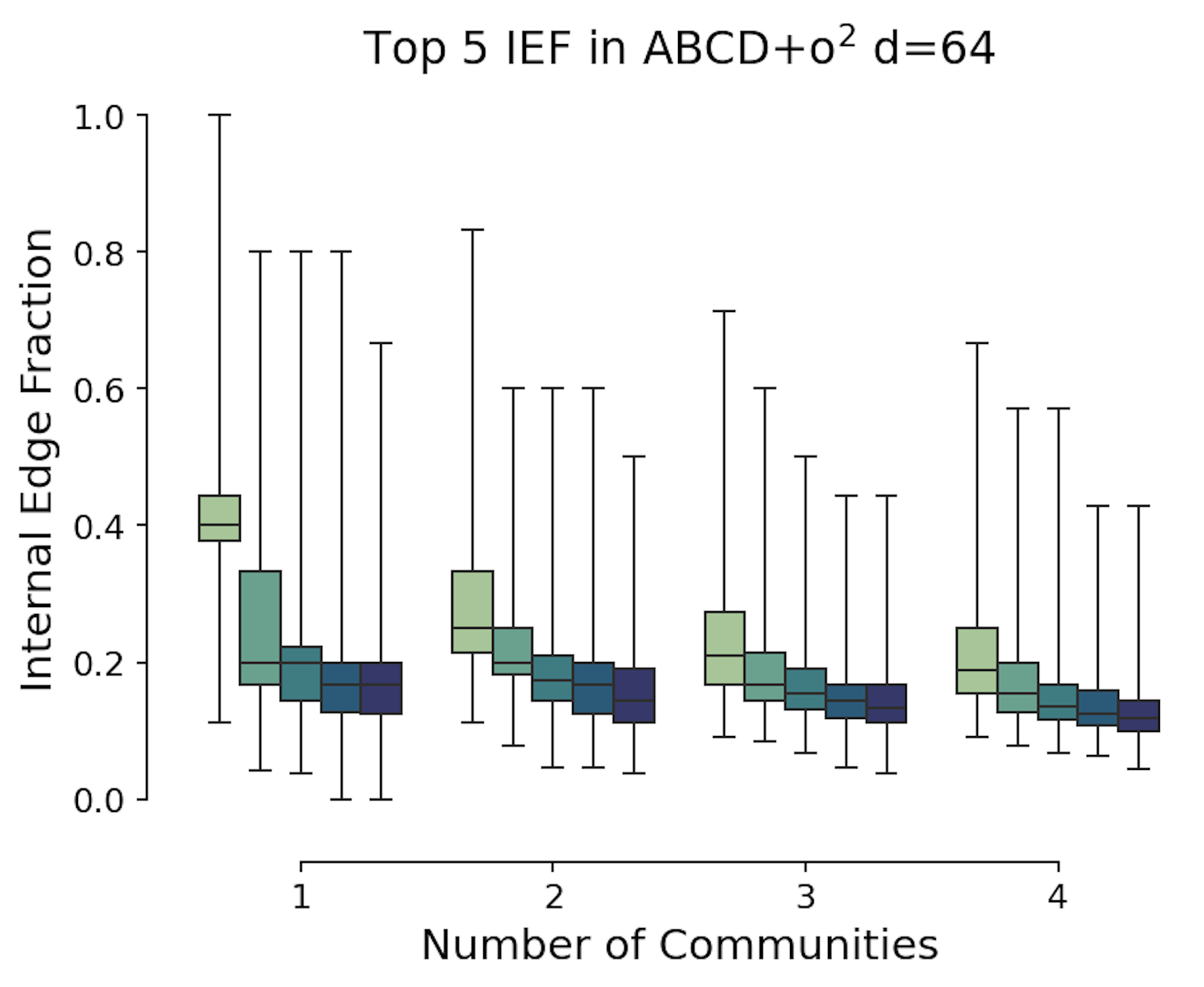}
    \caption{The top 5 IEF values in decreasing order, grouped by the number of communities the node belongs to, for YouTube and YouTube-like \ABCDoo graphs.}
    \label{fig:b3}
\end{figure}

Figures~\ref{fig:b1},~\ref{fig:b2} and~\ref{fig:b3}, together with the previously shown Figure~\ref{fig:dblp-ief}, present the full experiment showing the range of the top five internal edge fractions, sorted in descending order, on the real networks (left) and the corresponding \ABCDoo\ graphs (right), binned by the number of communities the node $v$ is in.

\end{document}